\nonstopmode
\documentclass{jfm}

\usepackage{graphicx}
\usepackage{newtxtext}
\usepackage{newtxmath}
\usepackage{natbib}
\usepackage{hyperref}
\usepackage{subcaption}

\usepackage{dakling}

\newcommand*\textcite{\citet}
\newcommand*\parencite{\citep}

\hypersetup{
    colorlinks = true,
    urlcolor   = blue,
    citecolor  = black,
}

\newcommand{\RomanNumeralCaps}[1]
\linenumbers

\title{Nonlinear optimals and their role in sustaining turbulence in channel flow}

\author{Dario Klingenberg \aff{1,}\aff{2}\corresp{\email{dario.klingenberg@zarm.uni-bremen.de}}
\and Rich R. Kerswell \aff{1} ,
 }

\affiliation{\aff{1}DAMTP, Centre for Mathematical Sciences, Wilberforce Road,
  Cambridge CB3 0WA, UK
\aff{2}ZARM, University of Bremen, Am Fallturm 2, 28359 Bremen, Germany
}

\begin{document}
\maketitle

\begin{abstract}
  We investigate the energy transfer from the mean profile to velocity
  fluctuations in channel flow by calculating nonlinear optimal disturbances,
  i.e.~the initial condition of a given finite energy that achieves the highest
possible energy growth during a given fixed time horizon.
  It is found that for a large range of time horizons and initial disturbance
energies, the nonlinear optimal exhibits streak spacing and amplitude consistent
with DNS at least at \(\retau = 180\), which suggests that they isolate the
relevant physical mechanisms that sustain turbulence.
  Moreover, the time horizon necessary for a nonlinear disturbance to
  outperform a linear optimal is consistent with previous DNS-based estimates
  using eddy turnover time, which offers a new perspective on how some turbulent
time scales are determined.
\end{abstract}



\section{Introduction}
\label{sec:introduction}

When trying to understand how turbulence is sustained in shear flow, an
essential part of the picture is the transfer of energy from the mean profile to
the fluctuations. In order to shed light on this, a popular approach has been to
linearise the Navier--Stokes equations around the mean profile and to study
the growth of linear optimal disturbances on top of this mean.
Some aspects of the dynamics observed in fully developed turbulence can be
explained in this way, for example streak formation along with the correct
streak spacing \citep{butler1993}, as well as streak breakdown
\citep{schoppa2002, hoepffner2005, cassinelli2017}.

This energy transfer is usually understood as a
self-sustaining process \citep{kim1971, jimenez1991, hamilton1995, waleffe1997,
  jimenez1999, farrell2012}.
The current prevailing picture is that of a two-stage linear
process \citep{schoppa2002, lozano-duran2021, kerswell2022}.
In a primary linear process, streaks form and grow due to transient
growth. If one subsequently changes the decomposition so as to include these
streaks in the base profile, a secondary linear process can describe how
secondary disturbances grow on top of this streaky field (again fuelled by
transient growth), ultimately causing the streaks to break down and restarting
the cycle \citep{hamilton1995, schoppa2002,hoepffner2005, lozano-duran2021}.

Even though breaking this process down into linear subprocesses offers great
practical advantages associated with the simplicity of linear theories in
general, it also introduces a number of questions that are difficult to answer
in a self-contained manner. For example, it remains unclear how to choose the
correct streak amplitude for the secondary linear process without appealing to
DNS results \parencite{kerswell2022}.

A natural generalisation, which also helps address these challenges, is to consider
the full nonlinear
optimisation problem. Thanks to the presence of the nonlinear term, the primary
and secondary linear processes can be coupled together, thus eliminating the
need for formulating a two-stage process.
Of course, it is unclear \textit{a priori} if the nonlinear optimal actually involves
such a two-stage process, or if completely different dynamics are more
effective.  More generally, the central question to be addressed in this work is
which aspects of the nonlinear optimal resemble the dynamics of real turbulence.

The motivation for studying this problem is twofold. First, a
better understanding of how turbulence is sustained could inform modelling and flow
control tasks. Second, \citet{malkus1956} formulated the seminal idea that the
mean profile actually observed in practice is in some way optimal in its ability
to sustain turbulence (in concrete terms, he asserted that the mean profile
should be marginally stable in a statistical sense). Using the methods available
at the time, he focused on long-term (linear) stability rather than short-term
transient energy growth as the main mechanism responsible for maintaining
turbulence and ultimately failed \citep[e.g.][]{reynolds1967} to formulate a suitable
definition of optimality. However, the general notion is still appealing today,
and we hope that the present study could contribute to a discussion about
reviving this idea.

Attempts to understand turbulence by linearising around a suitable base flow
have been around for a long time \citep{malkus1956, reynolds1967}.
However, it was not until \citet{farrell1988}
that transient growth was popularised as an important linear mechanism for
describing energy transfer from the mean profile to the fluctuations.
In a seminal work, \citet{schoppa2002} furthermore showed that when linearising
around a streaky base flow, it is again transient growth that can best describe
energy transfer from the streaky mean field to the disturbances, thus giving
rise to the two-step picture described above.
Whereas the primary linear process is fairly well-understood, much debate has
surrounded the question of which linear process can best describe the secondary
process
\citep{waleffe1997, reddy1998, schoppa2002, hoepffner2005, farrell2012}.
Recently, however, in an extensive cause-and-effect study,
\textcite{lozano-duran2021} were able to confirm \posscite{schoppa2002}
assertion that transient growth is more important than other proposed mechanisms
such as linear instability or parametric instability, at least at the relatively
low Reynolds number of \(\retau = 180\).

By comparison, the nonlinear optimisation problem has received much less
attention. For one, nonlinear optimisation relying on solving the full
three-dimensional Navier--Stokes equations only became computationally feasible
during the 2010s \citep{kerswell2010, cherubini2010, cherubini2011,
  monokrousos2011, pringle2012, farano2016}.
However, even today, the computational effort remains high, leading to
relatively few works applying nonlinear optimisation to the turbulent setting
\citep{cherubini2013, farano2015, farano2017}.
Notably, however, these works employ a different formulation than used here,
rendering these works unsuitable to answer the questions we are presently
concerned with. This is discussed in greater detail in
\cref{sec:problem-formulation}.

Including the nonlinear term in the optimisation problem can lead to
fundamentally different optimals. Whereas linear optimals are global in
structure and only include a single streamwise and spanwise wave number mode,
nonlinear optimals combine multiple wave numbers and, under some circumstances,
can be highly localised in space.
Moreover, they can achieve much higher growth than linear optimals,
often by coupling together multiple linear mechanisms \parencite{pringle2012,
  kerswell2018}.
An important point to note is that unlike in the transition
problem, the disturbances relevant to the turbulent problem are not necessarily
small, which makes it more difficult to justify dropping the nonlinear term .
Even though, as discussed in \textcite{lozano-duran2021}, there are still good
reasons to consider linear theories, it does imply that the evolution of the
linear optimal is in practice influenced by the nonlinear term, which, at high
energies, may actually limit the gains it can produce.
As a result, nonlinear optimals are likely more practically relevant while at
the same time also offering deeper insights into the relevant dynamics of
turbulence.

This paper is organised as follows: In \cref{sec:problem-formulation}, we
discuss the problem formulation on a theoretical level, and
\cref{sec:computational-method} describes the numerical implementation. Results
are shown and discussed in \cref{sec:results}, and \cref{sec:conclusion}
concludes.

\section{Problem formulation}
\label{sec:problem-formulation}

In the present work, we consider the same problem as \citet{lozano-duran2021},
i.e.\ incompressible Newtonian channel flow at friction Reynolds number
\({\retau = \utau h / \nu = 180}\), where \(\utau\) is the wall friction velocity,
\(h\) the channel half-height and \(\nu\) the kinematic viscosity.  The problem is
governed by the Navier--Stokes equations
\begin{align}
\label{eq:nse}
  \nc{} &\coloneq \pd{\utot}{t}
            + \utot \cdot \nabla \utot
            + \nabla \ptot
          + \vv{f}
            - \frac{1}{\retau} \nabla^{2}{\utot} = 0, \\
\label{eq:nsekonti}
 \mc{} &\coloneq \nabla \cdot \utot{} = 0,
\end{align}
where \(\utot\) and \(\ptot\) are the dimensionless flow velocity and pressure,
respectively, and \(\vv{f}\) is an external volume forcing term, the reasons for
including which will become apparent later.
The coordinate system is oriented such that \(x\) points in streamwise
direction, \(y\) in wall-normal direction and \(z\) in spanwise direction.
We decompose velocity and pressure into a base and a disturbance part,
\begin{equation}
\label{eq:decomp}
  \utot = \ubase + \upert, \quad
  \ptot = \pbase + \ppert,
\end{equation}
where capital letters denote the base and tilde denotes the disturbance fields.
In this study, the base is obtained by temporal averaging an \textit{a priori} DNS
run. 
Since it is calculated in advance, \(\ubase\) can be thought of as a fixed
parameter entering the problem.
In the following, we set \(- \nabla \pbase = \vv{\hat{x}}\), with \(\vv{\hat{x}}\)
denoting the unit vector in \(x\)-direction.
Note that the streamwise-spanwise (but not temporal) average of \(\upert\),
which we denote \(\avg{\upert}{xz}\), is allowed to be non-zero.  Here, angular
brackets denote averaging across the directions in the index, so, for example,
\begin{equation}
  \label{eq:6}
  \avg{\upert}{t xz} = \frac{1}{T L_{x} L_{z}} \int_{0}^{T}\int_{0}^{L_{x}}\int_{0}^{L_{z}} \upert \, dz dx dt.
\end{equation}
This ensures consistency because \(\upert\) is able to change the
streamwise-spanwise averaged velocity \(\avg{\utot}{xz}\) as the disturbance
grows on the short time scales we are presently interested in, even though
\(\ubase\) remains fixed. However, in the limit of very long time horizons, the
temporal average \(\avg{\upert}{t}\) vanishes.
Inserting \cref{eq:decomp} into \cref{eq:nsekonti,eq:nse} yields
\begin{align}
  \nonumber
\nc{} \coloneq
&\pd{\ubase}{t}
            + \ubase \cdot \nabla \ubase
          - \vv{\hat{x}}
          + \vv{f}
            - \frac{1}{\retau} \nabla^{2}{\ubase}
 \\
\label{eq:nsebase}
  +&\pd{\upert}{t}
            + \ubase \cdot \nabla \upert
            + \upert \cdot \nabla \ubase
            + \upert \cdot \nabla \upert
            + \nabla \ppert
            - \frac{1}{\retau} \nabla^{2}{\upert}
= 0, \\
\label{eq:nsekontibase}
  \mc{} \coloneq
           &\nabla \cdot \ubase{}
 + \nabla \cdot \upert{} = 0.
\end{align}
We now set
\begin{equation}
  \label{eq:5}
  \vv{f} = \vv{\hat{x}} + \frac{1}{\retau} \nabla^{2}{\ubase}.
\end{equation}
Physically, this forcing term represents the background turbulence fluctuations
maintaining \(\ubase\). It is important to note that these background
fluctuations are separate from the disturbances \(\upert\) in our present
model. This point is further discussed below.
Together with \(\pdi{\ubase}{t} = 0\) due to statistical stationarity and
\(\ubase \cdot \nabla \ubase = \nabla \cdot \ubase = 0\) in channel flow, this yields an equation
system for the disturbances which reads
\begin{align}
\label{eq:nsepertnoreystress}
  \ncf{} &\coloneq \pd{\upert}{t}
            + \ubase \cdot \nabla \upert
            + \upert \cdot \nabla \ubase
            + \upert \cdot \nabla \upert
            + \nabla \ppert
            - \frac{1}{\retau} \nabla^{2}{\upert} = 0,
  \\
\label{eq:nsepertkonti}
  \mcf{} &\coloneq \nabla \cdot \upert{} = 0.
\end{align}
For a given base profile \(\ubase\), time horizon \(T\) and initial condition
\(\upert(\xv, 0)\), \labelcref{eq:nsepertkonti,eq:nsepertnoreystress} can be
integrated forward in time to yield \(\upert(\xv, T)\), and, in particular, the
final disturbance energy
\(e_{T} \coloneq {1}/{(2 V)} \volint{\norm{\upert(\xv, T)}^{2}}\).
Normalising \(e_{T}\) by the initial energy \(e_{0}\) yields the energy gain
\(G \coloneq {e_{T}}/{e_{0}}\).
By further fixing the initial energy \(e_{0}\), one can ask which
initial condition \(\upert(\xv, 0)\) experiences the highest energy gain \(G\)
by the end of the time horizon \(T\).
This question naturally leads to an optimisation problem which has been studied
extensively in previous works \citep{kerswell2010, cherubini2010, cherubini2011,
  monokrousos2011, cherubini2013, farano2017}, and which can be solved using
adjoint looping.  Details of the solution algorithm are given in
\cref{sec:computational-method}.
In this study, only \cref{eq:nsepertnoreystress,eq:nsepertkonti} are actually
solved numerically. The base velocity is imposed to be the time-averaged mean
profile obtained from a previous DNS \citep{kerswell2022}.

As mentioned above, \(\avg{\upert}{xz}\) can be nonzero, and, as a consequence,
\(\avg{\utot}{xz}=\ubase+\avg{\upert}{xz}  \neq \ubase\).  Thus, it might appear like the initial
disturbance \(\upert(\xv, 0)\) is free to adjust \(\avg{\utot(\xv, 0)}{xz}\) to
any arbitrary profile, in particular one that might be more suitable for
generating disturbance energy growth than \(\ubase\).  
However, the initial energies considered here are  $\lesssim 0.02\%$ of the base
energy (except for one run at $0.04\%$ in \S 4.3) and so this is not an issue.

The formulation given by \cref{eq:nsepertnoreystress,eq:nsepertkonti} is
intended to be the simplest possible nonlinear model that could be used to
investigate disturbance growth in turbulent flows.
We note that in the literature, an alternative formulation exist
\citep[e.g.][]{farano2017}, which differs from our work in that no background
turbulence fluctuations are presumed.
This, in turn, implies a feedback mechanism: Since in that model, the
disturbances are responsible for maintaining the base flow \(\ubase\), they are
constrained to be in an order of magnitude that would be necessary for
maintaining the mean flow in a DNS.  For the questions to be investigated in the
present work, this constrained behaviour of the disturbances would not make
sense, as we are interested in how the disturbances grow on top of the base
profile, drawing energy only from it.

Some authors considering the linear optimisation problem \citep{reynolds1972,
pujals2009, cossu2017}, have advocated for including an eddy viscosity in the
momentum equation for the disturbances, i.e.\ \labelcref{eq:nsepertnoreystress}.
%
This introduces an ad-hoc assumption that the eddy viscosity felt by the mean
profile is also the same as would be felt by the fluctuations.
In order to keep the model as simple as possible, we decide not to include an eddy
viscosity in this study, though its effect could be an interesting topic for
future work.

To formulate the optimisation algorithm, we follow \citet{kerswell2018} and
consider the Lagrangian
\begin{align}
\label{eq:lagrangian}
  \nonumber
  \lagrangian &= \volint{\frac{1}{2} \norm{\upert(\xv, T)}^{2}}
  + \lambda \left[ \volint{\frac{1}{2} \norm{\upert(\xv, 0)}^{2}} - e_{0} \right]
  \\
  &+ \timeint {\volint{\uadj(\xv, t) \cdot \ncf{}}}
  + \timeint {\volint{\padj(\xv, t) \mcf{}}},
\end{align}
where \(\lambda, \uadj\) and \(\padj\) are Lagrange multipliers.
Setting the first variation of \(\lagrangian\) to zero leads, after some
algebra, to the further conditions for the Lagrange multiplier fields that
\begin{align}
  \label{eq:comp}
  \upert(\xv, T) + \uadj(\xv, T) &= 0,
  \\
  \label{eq:adj}
\pd{\uadj}{t}
            + (\ubase + \upert) \cdot \nabla \uadj
            - \uadj \cdot \transpose{(\nabla (\ubase + \upert))}
            + \nabla \padj
            + \frac{1}{\retau} \nabla^{2}{\uadj} &= 0,
  \\
  \label{eq:adjkonti}
  \nabla \cdot \uadj &= 0,
\end{align}
provided that the boundary conditions for \(\uadj\) are the same as for \(\upert\).
Furthermore,
\begin{equation}
  \label{eq:lagrgrad}
  \vd{\lagrangian}{\upert(\xv, 0)} = \lambda \upert(\xv, 0) - \uadj(\xv, 0) = 0.
\end{equation}
Note that \cref{eq:adjkonti,eq:adj} form the adjoint Navier--Stokes equations,
and the Lagrange multipliers \(\uadj\) and \(\padj\) can be interpreted as
adjoint velocity and pressure (divided by the density), respectively.
This set of equations naturally gives rise to a looping algorithm to obtain
the optimal initial condition \(\upert[\text{opt}](\xv, 0)\)  that generates the
highest energy gain \(G\) at the end of the
time horizon \(T\): First, make an initial guess for
\(\upert[\text{opt}](\xv, 0)\), which we call \(\upert(\xv, 0)\), and
evolve it forward in time using the Navier--Stokes equations
\labelcref{eq:nsepertkonti,eq:nsepertnoreystress} until time \(t=T\).
Second, obtain the initial condition for the adjoint system \(\uadj(\xv, T)\)
using \cref{eq:comp} and integrate \cref{eq:adjkonti,eq:adj} backward in time
from \(t=T\) to \(t=0\). Third, now that the final value of the adjoint
variable \(\uadj(\xv, 0)\) is known, calculate
\(\vdi{\lagrangian}{\upert(\xv, 0)}\) using \cref{eq:lagrgrad} and use this
gradient to improve the current guess \(\upert(\xv, 0)\). Repeat this loop until
some convergence criterion is reached.
More technical details of this algorithm are discussed in
\cref{sec:computational-method}.

\section{Computational method}
\label{sec:computational-method}

A special-purpose solver has been implemented in the programming language Python
to integrate \labelcref{eq:nsepertkonti,eq:nsepertnoreystress} forward in time
and to integrate the adjoint system \labelcref{eq:adjkonti,eq:adj}
backward in time.
This solver utilises the library JAX \citep{jax2018github},
which makes the code fully differentiable (although this is not used much in the
present work) and provides GPU-acceleration, thus delivering a crucial speed-up
and making the investigation feasible.
The source code of the solver is available at
\url{https://github.com/dakling/jax-spectral-dns}.

Following \citet{kim1987} and \citet{hoyas2021}, a velocity-vorticity formulation is
employed. Spatial directions are discretised using a pseudospectral method, with
Fourier modes in periodic directions and Chebyshev modes in the wall-normal
direction.  A third-order Runge--Kutta method derived in \citet{spalart1991} is
used for time discretisation.
De-aliasing using the \(3/2\)-rule \parencite{orszag1972} is done in the Fourier
directions (but not in Chebyshev directions for performance reasons), though
this is probably unnecessary at the employed resolution.  The nonlinear terms
are formulated in skew-symmetric form for increased numerical stability
\citep{gibson2014}.

As is further discussed in \cref{sec:infl-doma-size}, we choose a computational
domain size of \(\pi \times 2 \times \pi\), which is bigger than the minimal channel unit
(\(0.6\pi \times 2 \times 0.3\pi\) at the considered Reynolds number) but still small enough
to keep the numerical effort reasonably low \citep{jimenez1991}.
This domain size should be enough to observe some localisation of the nonlinear
optimal.
Following \citet{kim1987}, the domain is discretised
using a resolution of at least \(48 \times 129 \times 80\), which corresponds to
\(\Delta x^{+} \approx 11.8, \Delta y^{+}_\text{max} \approx 4.2\) and \(\Delta z^{+} \approx 7.1\), though some
calculations were done using a higher resolution to ensure that it does not make
a difference.
In order to make full use of JAX's GPU acceleration, the time step size should
be set \textit{a priori}, as JAX is better able to optimise loops of fixed size compared
to loops of variable size. Taking into account the base velocity and the
expected magnitude of the disturbances (which is either roughly known from the
previous looping iteration, or can be estimated depending on the nature of the
initial guess), the time step is chosen such that the CFL number is kept below
\(0.6\) with a safety factor of \(0.9\). Since the time stepping algorithm has
been reported to be stable up to \(\text{CFL}=0.7\) \citep{hoyas2021}, this
choice is conservative, and indeed no issues with time stepping instabilities
occurred with this setting.
Thanks to the GPU acceleration provided by JAX, the duration of a single DNS
forward run is in the order of a few minutes (compared to a few hours when
running on a single CPU), rendering the looping approach computationally
feasible.

More important than runtime, a major limiting factor for the present looping
algorithm is memory. Note that \cref{eq:adj} contains \(\upert\), i.e.\ the full
velocity spacetime history of the forward run. Depending on the details of the
setup, in our present investigation, the size of this variable is in the order
of magnitude between ten and 100 GB, which, given that this is just a single
variable, would cause out-of-memory issues on GPUs for all but the smallest
cases considered here if it needed to be saved entirely.  However,  memory can
be traded for runtime efficiency by employing a strategy known as checkpointing
\citep{berggren1998, griewank2000, hinze2005}.
The idea is that rather than storing the entire history of the forward run, only
some snapshots, or checkpoints, of the forward calculation are saved, serving as
initial conditions for recomputing intervals as needed during the adjoint
calculation.  The strategy employed in this work is fairly simple: snapshots are
saved at equally spaced time intervals. The number of time intervals is
chosen such that the number of snapshots is roughly equal to the number of
time steps between two snapshots. This reduces the memory requirements from \(N\)
to \(2 \sqrt{N}\), where \(N\) characterises the size of the spacetime history
of the velocity field, while increasing the runtime of the adjoint calculation
by a factor of two.

The effectiveness and efficiency of the looping approach crucially depends on
the employed gradient-based optimisation algorithm.
In the present work, we follow \textcite{cherubini2011} and use a conjugate
gradient method, although we determine the step size adaptively using a two-way
backtracking line search \parencite{truong2020}.
This line search is similar to a regular backtracking line search, but the step
size from the last iteration is used as an initial guess for the new step size,
thus reducing the number of necessary function evaluations (which are quite
expensive as they involve the forward run of the DNS).
If the step size is found to be sufficiently small, it is also periodically
tested if the step size can be increased, which is not necessary in classical
backtracking line search.
Note that the formula for the gradient \labelcref{eq:lagrgrad} contains the
Lagrange multiplier \(\lambda\), which ensures that the initial energy
constraint is fulfilled. Its appropriate value depends on the step size and
direction taken, so the gradient itself depends on the parameters of the
conjugate descent and must be recomputed appropriately. For a given set of
descent parameters, \(\lambda\) is then obtained using a standard Newton
method. To ensure that numerical issues with determining \(\lambda\) can never
lead to a violation of the initial energy constraint, the resulting initial
condition is also appropriately rescaled after each optimisation iteration.
The algorithm is documented in detail in \cref{sec:grad-desc-algor}.

Another critical question concerns constructing an appropriate initial guess.
Inconveniently, for many of the parameters considered here, it appears that the
linear optimal remains a local optimum, so that it (or slightly perturbed
versions of it) are unsuitable as an initial guess.
Instead, in this work, we rely on introducing randomness by running the
optimisation loop with a deliberately large step size, effectively forcing the
algorithm to sample the space of flow fields in a chaotic manner.
Another strategy is to use a random snapshot from a DNS as an initial condition.
Generally, it should be noted that even though nonlinear optimal outperform
linear ones even for relatively short time horizons and low initial energies, to
find them initially, longer time horizons and higher initial energies are
usually necessary.  Once an initial condition that gives rise to a nonlinear
optimal is found, this resulting nonlinear optimal can in turn also be used as
an initial conditions for calculations with different parameters.
Most of the parameter combinations -- including all that we focus the following
discussion on -- were also rerun from different initial
conditions to ensure that the same optimal was reached, although, since this is
a nonlinear optimisation problem, there is no rigorous way to ensure that a
global optimum has indeed been found.
Nevertheless, the conclusions made in the following are quite robust to changes
in the computational setup, including the domain size, as is further discussed
in \cref{sec:infl-doma-size}.

Since the code written for this study supports automatic differentiation, it
would also be possible to obtain the necessary gradient information without
solving the adjoint system. However, it has been found that using the classical
adjoint-based approach performs better than automatic differentiation, both in
terms of memory and runtime. In particular, the ability to more easily control
the trade-off between memory usage and runtime through checkpointing makes the
adjoint approach preferable since we are currently already close to the limits
of available GPU-memory on an NVIDIA A100 card with 40 GB memory.
Moreover, the gradients obtained from the adjoint method are fairly smooth,
whereas in gradients obtained from automatic differentiation, the discretisation
error manifests itself in the form of noise, so that some processing of these
gradients would likely be necessary.
Automatic differentiation does have the advantage of requiring less development
time and thus being more readily applied to new problems, so that this feature
will likely be valuable in future studies.

\section{Results}
\label{sec:results}

As discussed in \cref{sec:problem-formulation}, the main parameters are the time
horizon \(T\) and the initial disturbance energy \(e_{0}\).
Unlike in the transition problem, it is not useful to look at very large time
horizons, as the timescale on which disturbances can grow is constrained by
large-scale disruptive turbulent events. This argument has been used by
\textcite{butler1993} to determine the timescale for linear optimal growth, and
the general idea also applies to the nonlinear case.
From previous investigations \parencite{lozano-duran2021, kerswell2022}, we can
estimate that the primary and the secondary
linear process both operate on a timescale of around
\(t_{p} = t_{s} = 0.35 h / \utau\),  and that bursting events are roughly
\(t_{\text{burst}} = 4 h / \utau\) apart.
This motivates an initial focus on the time horizon
\(T = 0.7 h/\utau = 2 t_{p}\), as this would be enough to see both the primary
and the secondary linear process.  This is addressed in
\cref{sec:inner-time-scale}.
To get a deeper understanding of the role of this parameter, we additionally
study times \(t_{p}, 1.5 t_{p}, 3 t_{p}, 4 t_{p}, 6 t_{p}\) and \(8 t_{p}\) in
\cref{sec:expl-time-energy}.

For any \(T\), as \(e_{0} \rightarrow 0\), the linear optimisation problem is recovered.
Consequently, for small \(e_{0}\), a quasilinear optimal, which closely
resembles the linear optimal, is obtained.
The linear optimal's distinguishing property is that it consists of only
a single wave number mode. This is because interactions between different
wave number modes would require nonlinear effects, in the absence of which the
optimisiation algorithm simply picks the wave number mode with the highest energy
gain for the given time horizon.
The quasilinear optimal is the weakly nonlinear extension of the linear optimal.
Since it is finite amplitude, nonlinear coupling between modes exists but is too
weak to make much of a difference compared to the linear optimal.
As \(e_{0}\) increases, there comes a point where a structurally completely
different nonlinear optimal arises.
For large \(e_{0}\), the final state of the disturbances may be well in the
turbulent regime, which, in the current formulation, usually leads the
optimisation algorithm to fail, as the final state dependency on the initial
condition is too sensitive and chaotic for the optimisation problem to remain
well-conditioned. The precise determination of this boundary is difficult and
not very important to the present study, but values up to
\(e_{0}/E_{0} = \expnumber{-4}\) were found to converge without issues even for
the longest time horizons considered here, where \(E_{0}\) denotes the energy of
the base flow.
Naturally, we are most interested in nonlinear optimals whose initial energy is
below this value.

\subsection{Optimal disturbances at \(T=0.7 h/\utau\)}
\label{sec:inner-time-scale}

\begin{figure}
  \centering
  \includegraphics[width=0.9\textwidth]{ 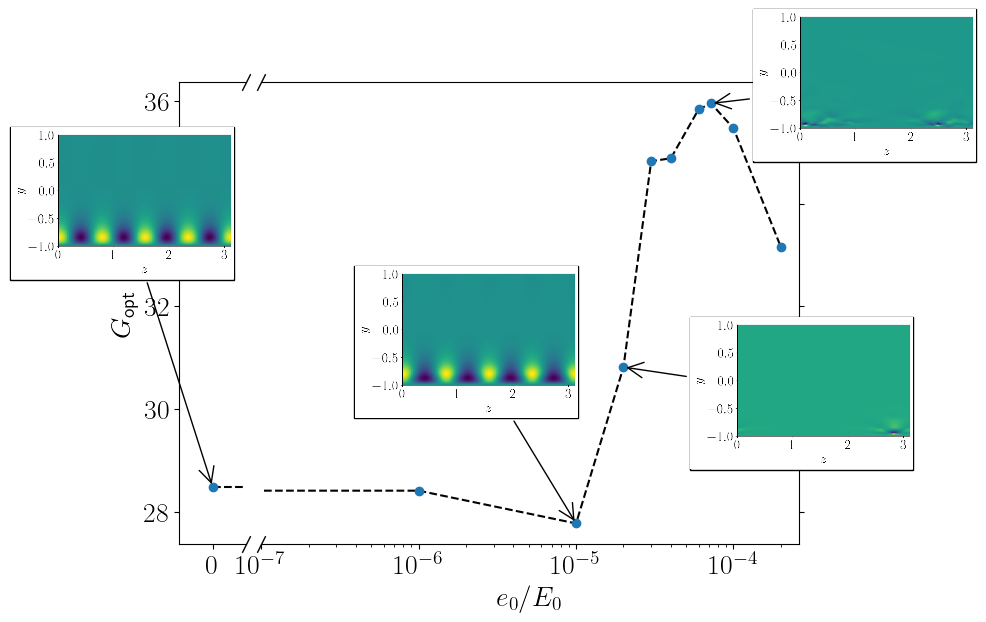 }
  \caption{Optimal gain over initial energy normalised by the energy of the base
flow \(E_{0}\). Note the truncated vertical axis. Plots of the initial condition
of the streamwise disturbance velocity for some of the optimals are also
included. For more details on these and their time evolution please refer to
\cref{fig:vel-2t0-lin,fig:vel-2t0-7eminus5}.}
  \label{fig:gain-over-energy-2t0}
\end{figure}

In this \namecref{sec:inner-time-scale}, we consider the time horizon
\(T = 0.7 h / \utau\).
The optimal gain that can be achieved as a function of initial energy \(e_{0}\)
is plotted in \cref{fig:gain-over-energy-2t0}.

\begin{figure}
  \centering
  \includegraphics[width=0.95\textwidth]{ 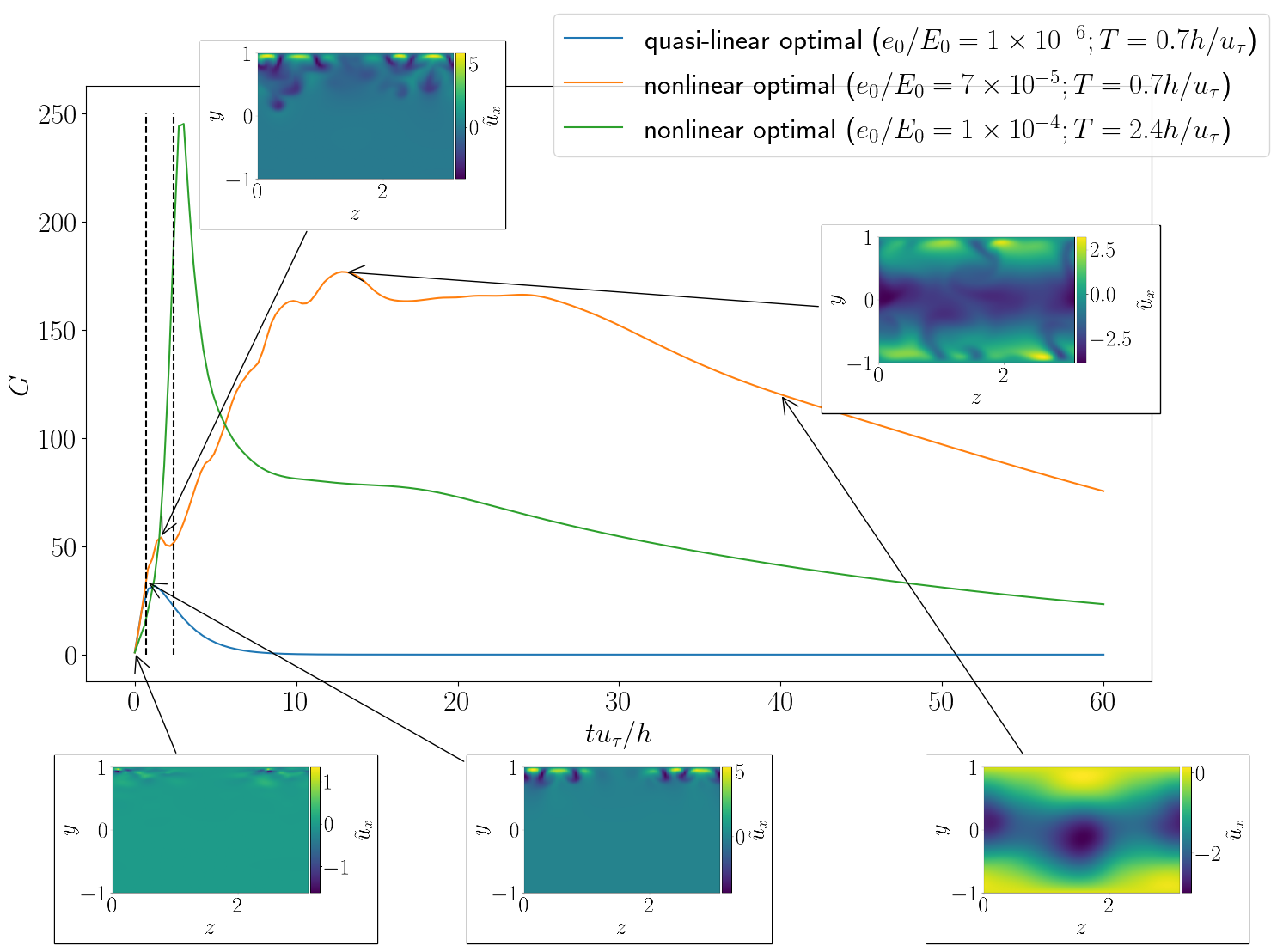 }
  \caption{Long-term evolution (\(t_{\text{final}} \utau/h = 60\))
    of the energy gain \(G\) of the quasilinear optimal
    (\(e_0/E_{0} = \expnumber{-6}, T \utau/h=0.7\), blue line), the nonlinear optimal
    (\(e_0/E_{0} = \expnumber[7.2]{-5}, T \utau/h=0.7\), orange line),
    as well as the nonlinear optimal (\(e_0/E_{0} = \expnumber[1]{-4},
    T \utau / h = 2.8\), green line).
    Vertical dashed lines indicate \(t \utau/h = 0.7\) and \(t \utau/h = 2.8\),
    respectively.
    Snapshots of the streamwise disturbance velocity are shown for the
(\(e_0/E_{0} = \expnumber[7.2]{-5}\), \(T \utau / h = 0.7\))-optimal (orange
line).
}
  \label{fig:energy-2t0-7eminus5}
\end{figure}

It is apparent that at \(e_{0}/E_{0} = \expnumber[2]{-5}\), a first nonlinear
optimal emerges. As the initial energy increases further, this optimal becomes
more and more efficient, until the highest gain among all \(e_{0}\) is found at
\(e_{0}/E_{0} = \expnumber[7.2]{-5}\). As \(e_{0}/E_{0}\) is increased beyond
this value, saturation seems to set in, and the achieved gains slowly diminish,
until around \(e_{0}/E_{0} = \expnumber[2]{-4}\), the optimisation problem
starts becoming ill-conditioned.
Interestingly, the difference between the highest nonlinear optimal gain
(\(34.82\)) and the linear gain (\(28.48\)) is not very high at this \(T\).
However, the long-term evolution (obtained by solving
\cref{eq:nsepertnoreystress,eq:nsepertkonti} with an optimal disturbance as the
initial condition, but for \(t_\text{final} \gg T\)) reveals a major difference:
whereas the gain
\(G(t) = e(t) / e_{0} = {1}/{(2V)} \volint{\norm{\upert(\xv, t)}^{2}} / e_{0}\) of the linear
optimal tends to reach its peak gain at or close to \(t = T\), the nonlinear
optimals keep growing significantly, often almost by another order of magnitude
(see \cref{fig:energy-2t0-7eminus5}).
This continued growth seems to be a consequence of localisation, i.e.\ it is mainly
due to the disturbance further spreading out in space rather than it growing in
amplitude.
The fact that a decay is observed for long times 
ensures that the long-term temporal average \(\avg{\upert}{t}\) vanishes.

\begin{figure}
  \centering
  \includegraphics[width=1.0\textwidth]{ 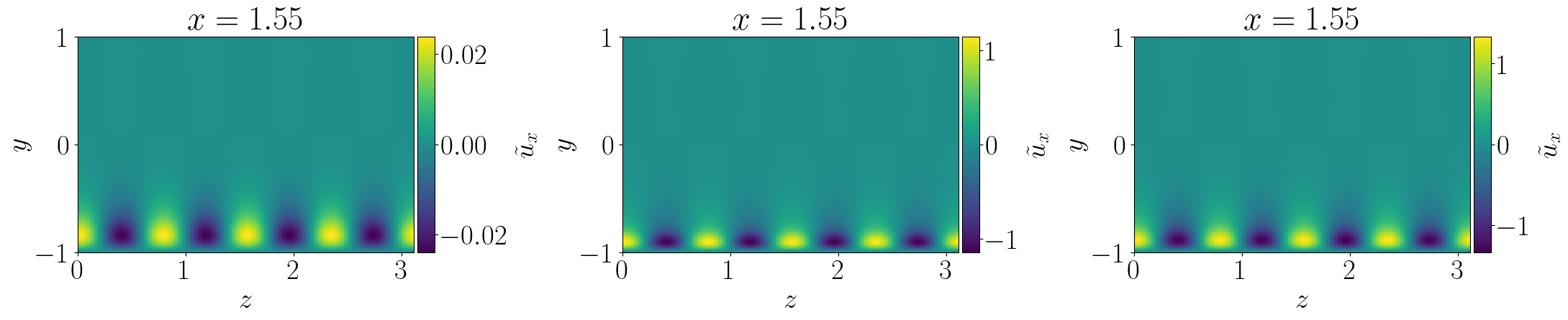 }
  \caption{Streamwise velocity evolution (\(t \utau/h = 0, 0.35, 0.7\)) of the linear optimal for \(T \utau/h=0.7\).}
  \label{fig:vel-2t0-lin}
\end{figure}

Next, we study the structure of the observed optimals to better assess their
relevance in fully developed turbulence, starting with the linear optimal.
Its time evolution is shown in \cref{fig:vel-2t0-lin}.
As is expected for a linear optimal, only a single wave number mode -- in this
case \([k_{x}, k_{z}] = [0, 8]\) -- is present, with the energy of all other modes
combined only making up around \(\expnumber{-10}\) times the total energy.
This wave number mode grows due to the well-documented transient growth
mechanism based on non-normal interaction of modes.
Here, \(k_{x} = 2\pi / \lambda_{x}\), where  \(\lambda_x\) stands for the
\(x\)-wavelength, and \(k_{z}\) is defined analogously.
Note that the streak spacing does not match what one would observe in real
turbulence. This is expected, however, because, following the argument of
\textcite{butler1993}, the considered time horizon is unrealistically long (by a
factor of two) for the primary linear process alone.

\begin{figure}
  \centering
  \includegraphics[width=1.0\textwidth]{ 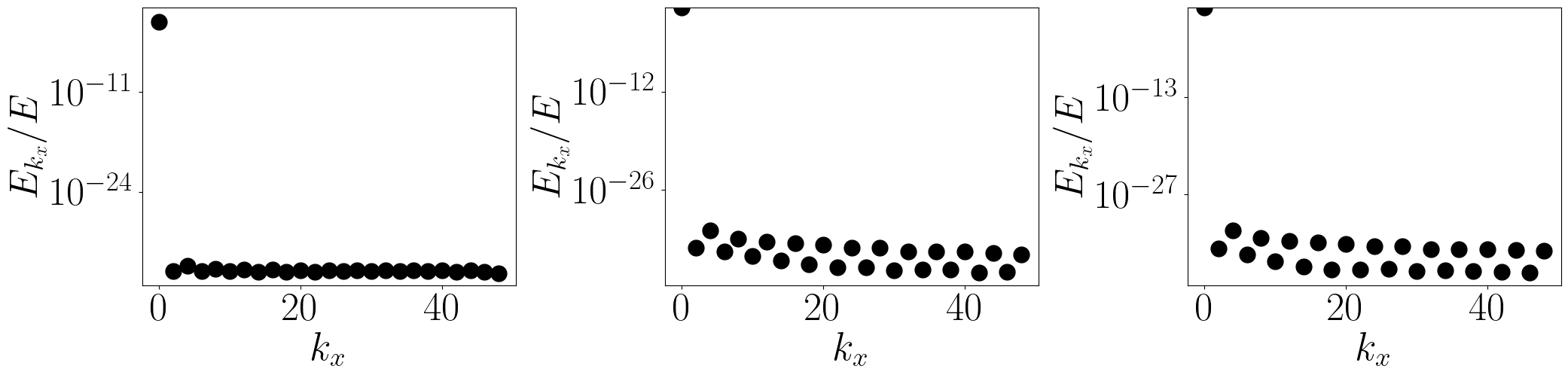 }
  \caption{Streamwise spectrum of the streamwise velocity at times
\(t \utau/h = 0, 0.35, 0.7\) of the quasilinear optimal
(\(e_{0}/E_{0} = \expnumber{-6}\)) for \(T \utau/h=0.7\). Essentially all of the
energy is contained in the \(k_{x} = 0\) mode.}
  \label{fig:vel-2t0-1eminus6-spectrum}
\end{figure}

Moving to finite but small values for \(e_{0}/E_{0}\), at \(\expnumber{-6}\) and
up to \(\expnumber{-5}\), a quasilinear optimal is found. Its velocity time
evolution (not shown) closely resembles the linear optimal
(\cref{fig:vel-2t0-lin}), though the effects of the nonlinear term are visible
in the combined energy of the non-dominant wave modes, which grows from around
\(\expnumber[3]{-4}\) to \(\expnumber[2.6]{-3}\) times the total energy.
This becomes apparent in the spectral decomposition of the streamwise velocity
shown in \cref{fig:vel-2t0-1eminus6-spectrum}.

\begin{figure}
  \centering
  \includegraphics[width=1.0\textwidth]{ 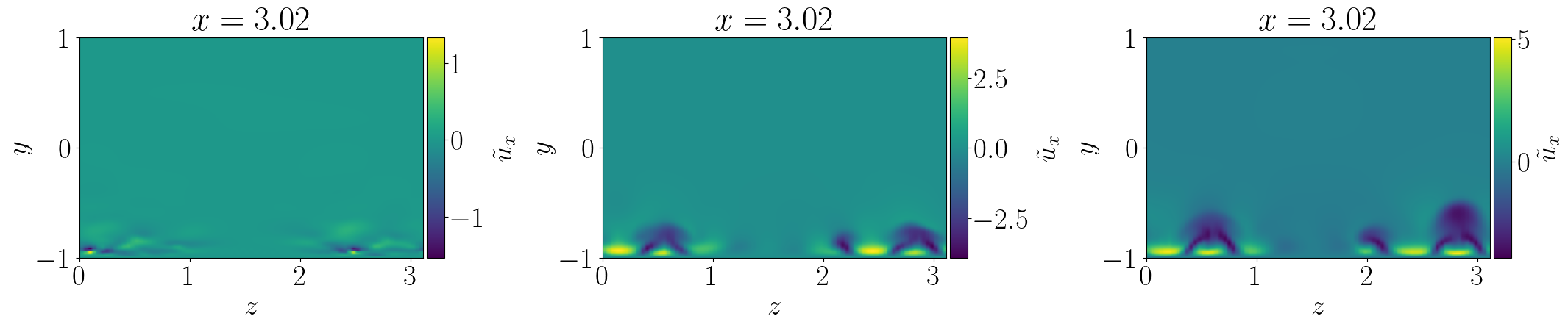 }
  \caption{Streamwise velocity evolution (\(t \utau/h = 0, 0.35, 0.7\)) of the
nonlinear optimal (\(e_{0} / E_{0} = \expnumber[7.2]{-5}\)) for \(T \utau/h=0.7\)
plotted in \(y\)-\(z\)-plane.}
  \label{fig:vel-2t0-7eminus5}
\end{figure}

\begin{figure}
  \centering
  \includegraphics[width=1.0\textwidth]{ 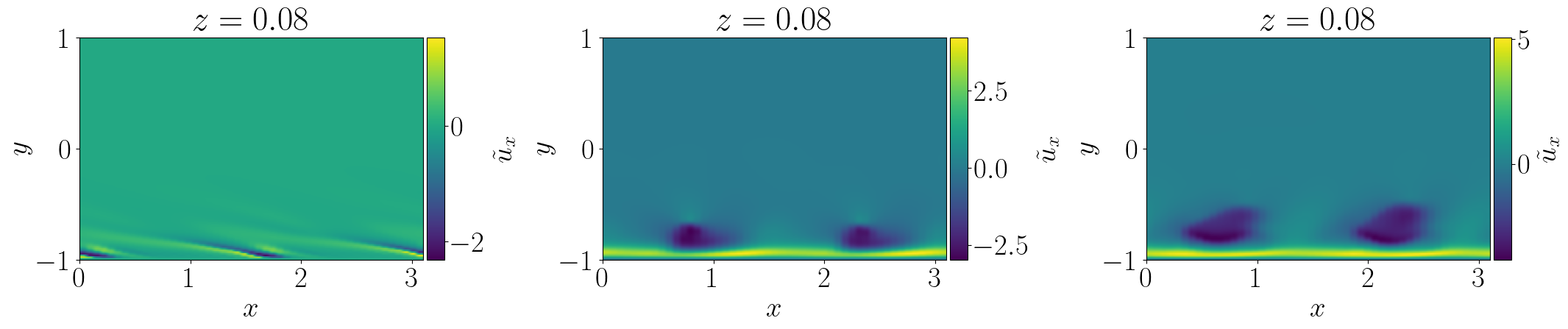 }
  \caption{Streamwise velocity evolution (\(t \utau/h = 0, 0.35, 0.7\)) of the
    nonlinear optimal (\(e_{0} / E_{0} = \expnumber[7.2]{-5}\)) for
    \(T \utau/h=0.7\) plotted in \(x\)-\(y\)-plane.}
  \label{fig:vel-2t0-7eminus5_xy_plane}
\end{figure}

Optimal gain is achieved by the nonlinear optimal with initial energy
\(e_{0}/E_{0} = \expnumber[7.2]{-5}\), and its time evolution is shown in
\cref{fig:vel-2t0-7eminus5,fig:vel-2t0-7eminus5_xy_plane}.
It is apparent that the shape of the initial condition is completely different
from the linear optimal. Most notably, no streaks are present, and the
disturbance is concentrated, or localised, in specific parts of the domain.
This localisation is most clear in spanwise direction, as there is no full
localisation in the streamwise direction (see \cref{fig:vel-2t0-7eminus5_xy_plane}).
Instead, in streamwise direction, there
seems to be a length scale between the two distinct peaks of the disturbance.
Indeed, when moving to a box twice as large in streamwise direction (i.e.\
\(2 \pi \times 2 \times \pi\)), four such
peaks can be identified (see \cref{sec:infl-doma-size}).
It is unclear how this length scale is determined, but we may speculate that it
is the most efficient spacing to quickly generate continuous streaks.  This may
also be related to the size of the minimal channel unit, which is roughly half
of the current domain in streamwise direction. Therefore, it is possible that
the current structure could be thought of as two minimal channel optimals
stitched together in streamwise direction.
Localisation is also visible in the wall-normal direction, with the initial
disturbance concentrated close to the wall where the base shear is high.
The fact that the entire disturbance is concentrated on one side and not
symmetrical can also be explained with localisation.
This is in contrast to the linear optimal, for which both the symmetrical and
the one-sided version yield the same gain, though we only show the one-sided
version here.
Localisation is a well-known phenomenon for nonlinear optimals
\parencite{kerswell2018}, and can be rationalised by the optimisation algorithm
trying to accommodate the initial energy constraint: instead of distributing the
disturbance energy evenly, it may make more sense to have high-amplitude
disturbances strategically placed in some small areas and then rely on
nonlinear mechanisms to spread out these disturbances over time.

\begin{figure}
  \centering
  \includegraphics[width=1.0\textwidth]{ 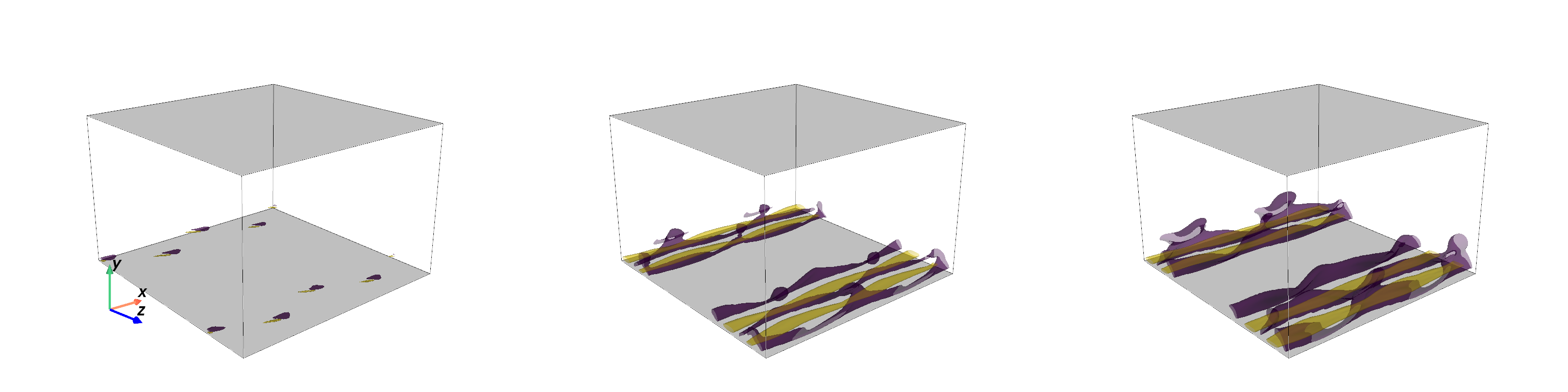 }
  \caption{Streamwise velocity evolution (\(t \utau/h = 0, 0.35, 0.7\)) of the
    nonlinear optimal (\(e_{0} / E_{0} = \expnumber[7.2]{-5}\)) for
    \(T \utau/h=0.7\). Yellow isocontours indicate 60 \% of the maximum value, and
    blue isocontours 60 \% of the minimum value.}
  \label{fig:vel-2t0-7eminus5_isosurfaces}
\end{figure}

\begin{figure}
  \centering
  \includegraphics[width=1.0\textwidth]{ 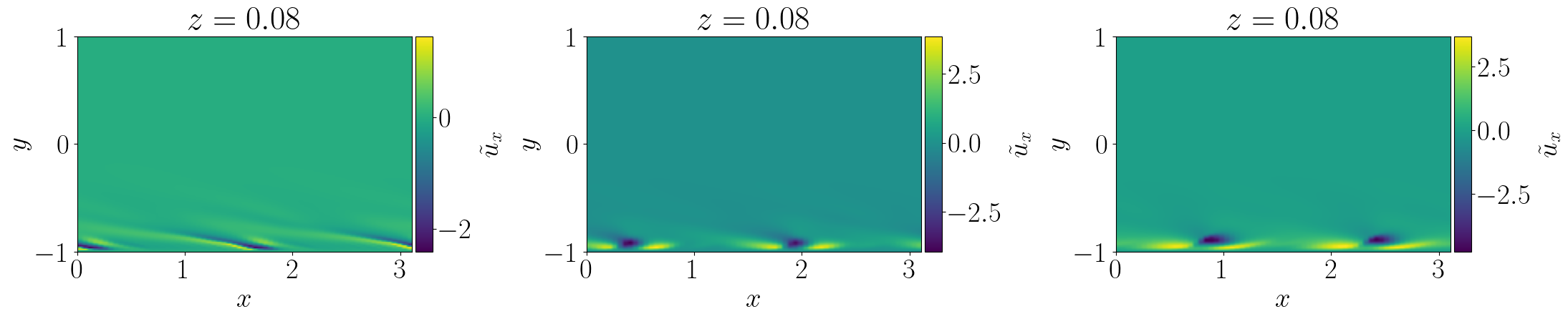 }
  \caption{Early part of the streamwise velocity evolution (\(t \utau/h = 0, 0.35, 0.7\)) of the
    nonlinear optimal (\(e_{0} / E_{0} = \expnumber[7.2]{-5}\)) for
    \(T \utau/h=0.7\) plotted in \(x\)-\(y\)-plane.}
  \label{fig:vel-2t0-7eminus5_xy_plane_short}
\end{figure}

\begin{figure}
  \centering
  \includegraphics[width=1.0\textwidth]{ 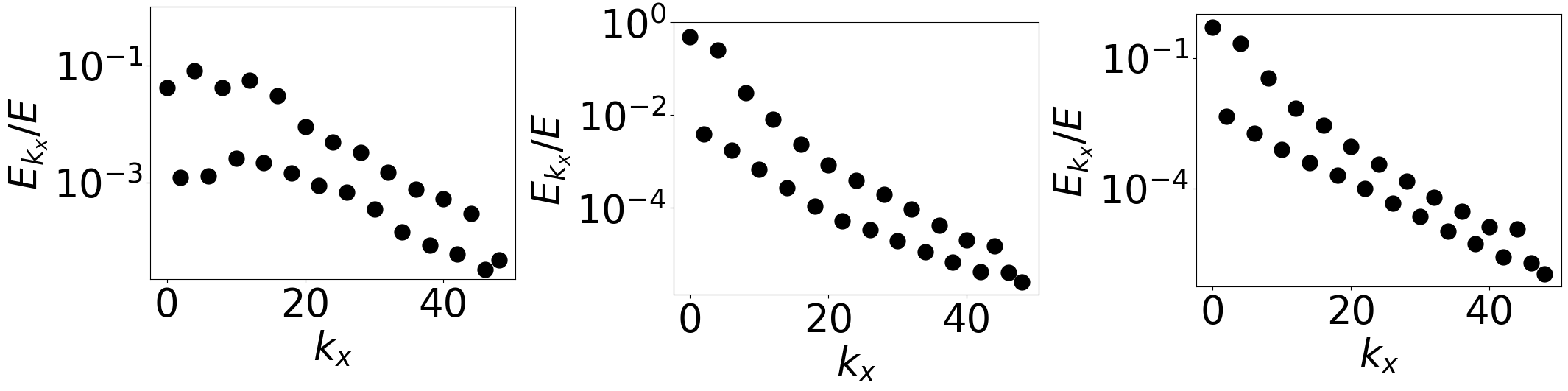 } 
  \caption{Streamwise spectrum of the streamwise velocity at times
\(t \utau/h = 0, 0.35, 0.7\) of the nonlinear optimal
(\(e_{0}/E_{0} = \expnumber[7]{-5}\)) for \(T \utau/h=0.7\).
Compared to \cref{fig:vel-2t0-1eminus6-spectrum}, the spectral dynamics are much
more complicated, with streaks (\(k_x = 0\)) not being dominant initially but
then containing most of the energy later.}
  \label{fig:vel-2t0-7eminus5-spectrum}
\end{figure}

\begin{figure}
  \centering
  \includegraphics[width=0.85\textwidth]{ 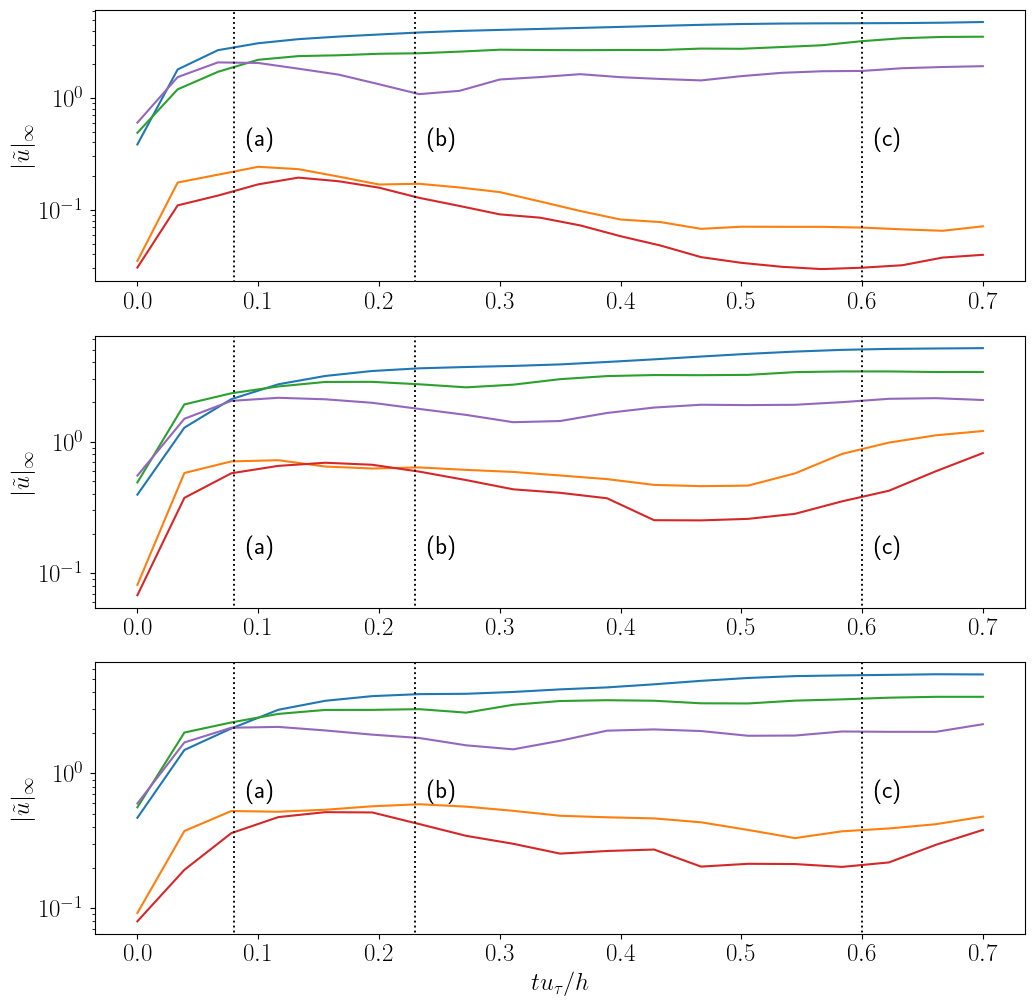 }
  \caption{Streamwise velocity amplitudes of \(x\)-wave number modes over time
    for different localised nonlinear optimals,
    \(e_{0} / E_{0} = \expnumber[3]{-5}\) (top),
    \(e_{0} / E_{0} = \expnumber[7.2]{-5}\) (middle) and
    \(e_{0} / E_{0} = \expnumber{-4}\) (bottom), all for \(T \utau/h=0.7\).
    Shown are wave numbers \(k_{x} = 0\), i.e.\  the streaks (blue);
    \(k_{x} = 2\) (orange); \(k_{x} = 4\) (green); \(k_{x} = 6\) (red) and
\(k_{x} = 8\) (purple).
    Despite the different initial disturbance energies, all exhibit a rapid early
    growth up to time (a), then a decay of the \(k_{x} \geq 6\)-modes
    until around time (b), and a final growth of these higher-order
    modes after time (c).
  }
  \label{fig:amps}
\end{figure}

The early part of the evolution is characterised by an absence of streaks (see
\cref{fig:vel-2t0-7eminus5-spectrum}, left tile), and
instead we observe tilted flow structures oriented such that they are amplified
by an Orr-type mechanism (this is most clearly observed in
\cref{fig:vel-2t0-7eminus5_xy_plane_short}).
Following this initial phase, these disturbances quickly organise into streaks
(see
\cref{fig:vel-2t0-7eminus5_xy_plane,fig:vel-2t0-7eminus5_isosurfaces,fig:vel-2t0-7eminus5-spectrum},
middle tile).
Note that the streaks that have formed at this point do exhibit correct streak
spacing and amplitude (see \cref{fig:vel-2t0-7eminus5}, middle tile, and the
discussion in \cref{sec:expl-time-energy}).  Unlike the streaks of the linear
optimal, the low-speed streaks do not remain straight as they evolve, but
meander and wrap around the high-speed streaks, as is revealed by
\cref{fig:vel-2t0-7eminus5_isosurfaces}.
Subsequently, these streaks grow due to the lift-up mechanism, and although no
clear streak breakdown can be observed, it is clear from
\cref{fig:amps} that second-order
disturbances grow on top of the streaks towards the end of the considered time
window (from time (c) onwards). Thus, it can be concluded that the
primary and, although not as pronounced, the secondary linear process are both
visible in the nonlinear optimum, though the way that they are coupled together
by the nonlinear term means that there is significant temporal overlap.  This
serves as a reminder that characterising them as neatly separated subprocesses
constitutes a major simplification.
Moreover, the streak growth itself is clearly not exclusively fuelled by
transient growth (although this likely plays a significant role), as can be most
clearly observed by the temporal evolution of the high-frequency modes
($k_x \geq 6$) between the times (a) and (b) shown in \cref{fig:amps}.
After an initial growth up to the time marked by vertical line (a), these modes
start decaying, indicating that they transfer energy to the streaks.

\begin{figure}
  \centering
\begin{subfigure}{0.3\textwidth}
  \centering
    \includegraphics[width=1.0\textwidth]{ 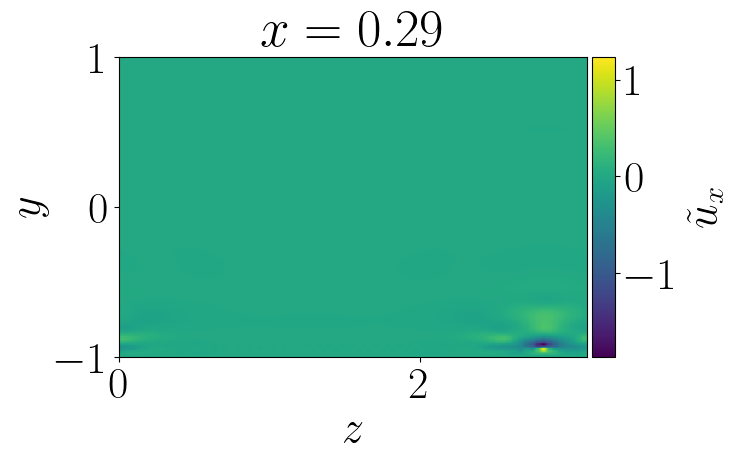 }
\end{subfigure}
\hspace{1em}
\begin{subfigure}{0.3\textwidth}
  \centering
    \includegraphics[width=1.0\textwidth]{ 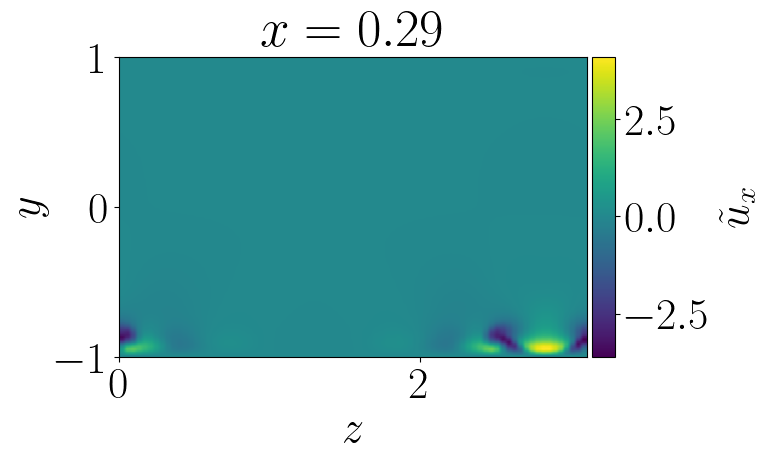 }
\end{subfigure}
\hspace{1em}
\begin{subfigure}{0.3\textwidth}
  \centering
    \includegraphics[width=1.0\textwidth]{ 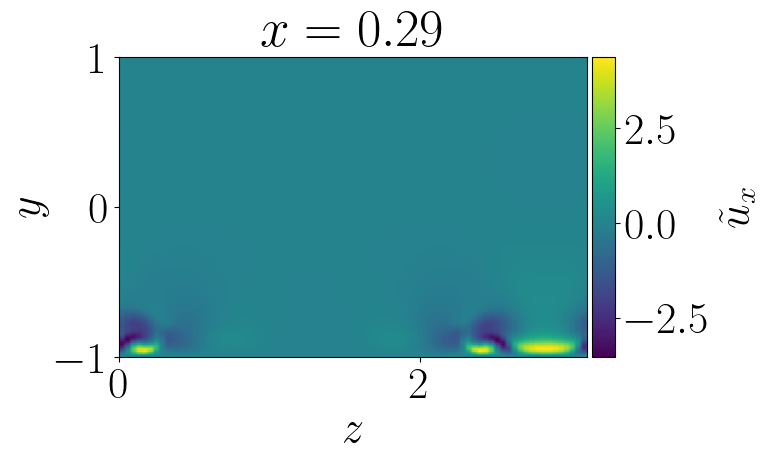 }
\end{subfigure}
\\
\begin{subfigure}{0.3\textwidth}
  \centering
    \includegraphics[width=1.0\textwidth]{  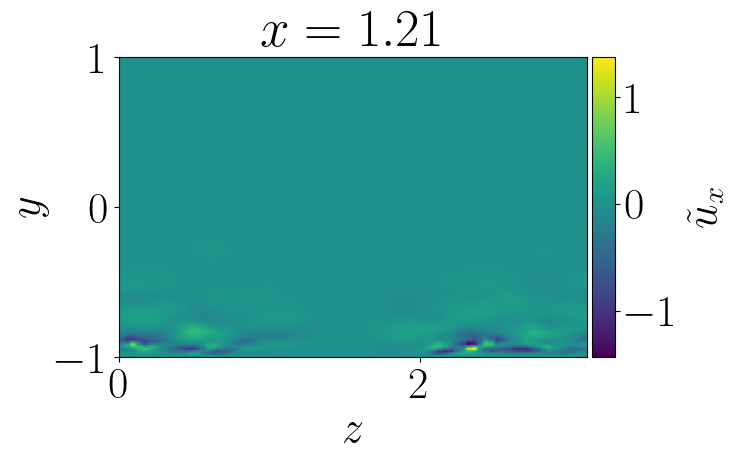 }
\end{subfigure}
\hspace{1em}
\begin{subfigure}{0.3\textwidth}
  \centering
    \includegraphics[width=1.0\textwidth]{  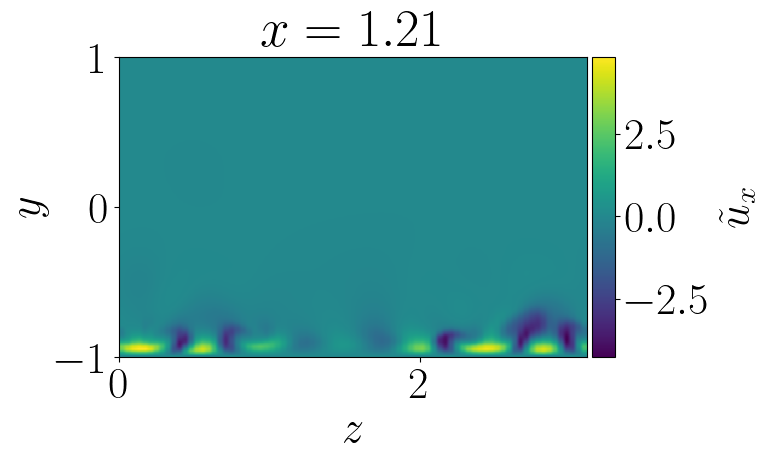 }
\end{subfigure}
\hspace{1.0em}
\begin{subfigure}{0.3\textwidth}
  \centering
    \includegraphics[width=0.91\textwidth]{  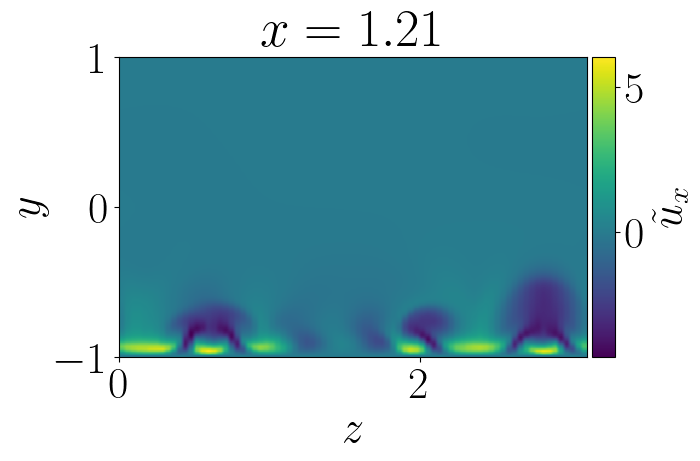 }
\end{subfigure}
  \caption{Comparison of nonlinear optimals and their evolution at different
  initial disturbance energies. Top row: \(e_{0}/E_{0} = \expnumber[3]{-5}\), bottom row:
  \(e_{0}/E_{0} = \expnumber{-4}\). Columns from left to right correspond to
  initial (\(t \utau/h=0\)), intermediate (\(t \utau/h=0.35\)) and final (\(t
  \utau/h=0.7\)) streamwise velocity field.}
  \label{fig:comp-nonlinear-optimals}
\end{figure}

Interestingly, the main features remain remarkably stable throughout the
nonlinear regime. Comparing the optimal at \(e_{0}/E_{0} = \expnumber[3]{-5}\) with
the one at \(e_{0}/E_{0} = \expnumber{-4}\)
(\cref{fig:comp-nonlinear-optimals}) reveals that only the localisation
in spanwise direction diminishes for higher energies (as one would expect), but
apart from this, the dynamics, including time scales and streak spacing, are
comparable. In particular, the minimum and
maximum values of the streamwise velocity disturbances are very similar, so it
can be concluded that the additional initial energy available for the
\(e_{0}/E_{0} = \expnumber{-4}\) case only serves to reduce localisation,
rather than amplify the magnitude of the disturbance.
The similarity of the underlying dynamics is also confirmed by the wave number
plot shown in \cref{fig:amps} and further
discussed in \cref{sec:expl-time-energy}.

\subsection{Exploring time-energy parameter space}
\label{sec:expl-time-energy}

\begin{figure}
  \centering
  \includegraphics[width=0.9\textwidth]{ 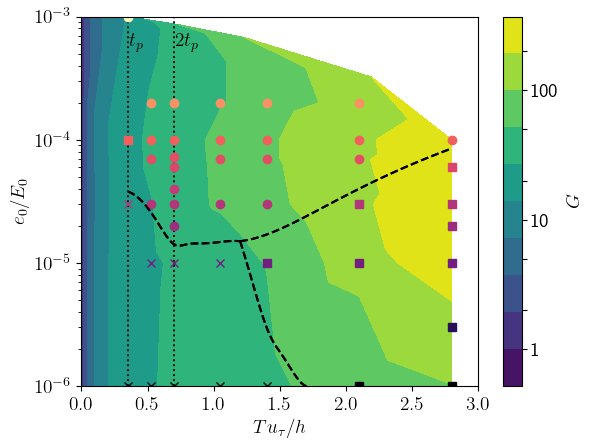 }
  \caption{$T$-$e_0$-parameter space. Crosses indicate the quasilinear regime,
    squares the nonlinear non-localised regime and circles the nonlinear localised
    regime.
    The dashed lines were added manually to aid visual distinction between the
    regimes.
    Furthermore, each dot or cross is also colour-coded according to its \(e_{0}\),
    which is redundant in this plot but allows establishing connections to
    \cref{fig:tinfnormparamspace,fig:tlambdazparamspace,fig:tstreakampparamspace}.
  }
  \label{fig:te0paramspace}
\end{figure}

It is clear that not any arbitrary parameter combination of \(T\) and \(e_{0}\)
will give rise to optimals that distil the relevant dynamics of fully-developed
turbulence.
Although rough estimates for the time and energy scales can be made based on
scaling arguments as discussed in \cref{sec:inner-time-scale}, it is important
to understand how the shape and dynamics of the optimals change in
\((T\),\(e_{0})\)-space.  In particular, we would like to investigate how sensitive
the dynamics are to changes in \(T\) and \(e_{0}\).
\Cref{fig:te0paramspace} shows the different regimes that can be identified
between \(T = 0.35 h/\utau\) and \(T = 2.8 h/\utau\), with the initial energy
chosen small enough to result in an end state that is not too chaotic
for the optimisation to converge.
The gain achieved by each optimal is also shown, with values between the points
obtained by interpolation.
For small \(e_{0}\), the quasilinear regime
is found, which we here identify by the most energetic wave number
mode of the initial optimal disturbance containing most (more than \(90 \%\),
though the exact value does not matter much as long as it is high) of the total
energy.
The boundary to the nonlinear regime has not been precisely
resolved for all \(T\) (unlike in the transition problem, this area is not of
primary interest here).
It is interesting to note, however, that at \(T \leq t_{p}\), the quasilinear
optimal found at low \(e_{0}\) produces higher gains than optimals found at
higher \(e_{0}\). This trend is reversed for larger \(T\), which implies that
the primary growth timescale \(t_{p}\) can also be thought of as the largest
time horizon for which the (quasi-) linear optimal outperforms the nonlinear one.

\begin{figure}
  \centering
  \includegraphics[width=1.0\textwidth]{ 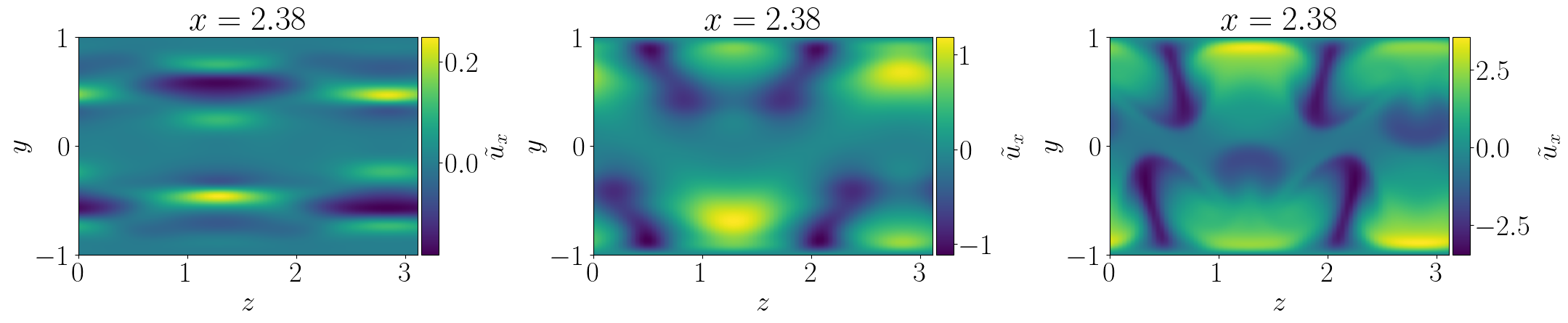 }
  \caption{Streamwise velocity evolution (\(t \utau/h = 0, 1.4, 2.8\)) of the optimal for \(T \utau/h=2.8\) and
    \(e_0/E_{0} = \expnumber[3]{-5}\).}
  \label{fig:vel-8t0-3eminus5}
\end{figure}

\begin{figure}
  \centering
  \includegraphics[width=0.5\textwidth]{ 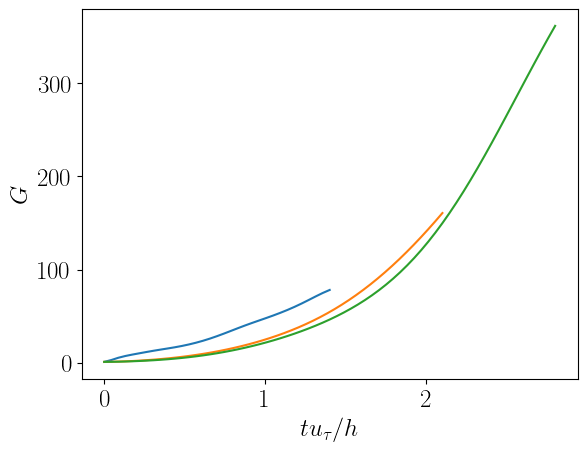 }
  \caption{Energy over time of the optimal for the nonlinear localised optimal
    at \(T \utau/h=1.4\) (blue) compared to the nonlinear non-localised optimals at
    \(T \utau/h=2.1\) (orange) and \(T \utau/h=2.8\) (green), all at
    \(e_0 / E_{0} = \expnumber[3]{-5}\).
    It is apparent that much of the energy growth of the non-localised optimals
    happens late, which explains why they are not observed for short time horizons.
  }
  \label{fig:energy-8t0-3eminus5}
\end{figure}

At large time horizons but small initial energies, a completely new nonlinear
optimal structure emerges (\cref{fig:vel-8t0-3eminus5}).
Interestingly, this structure is not localised but global, and, as can be seen in
\cref{fig:energy-8t0-3eminus5},
much of the energy growth happens in the final part of the calculation, which
explains why this structure is not optimal when considering shorter time
scales.
We distinguish this regime from the nonlinear localised regime by checking that
the lowest quarter of wave number modes of the initial optimal disturbance
contain almost all (\(>99.999\%\)) of its energy.
However, as is apparent visually and will be discussed in more detail later, the
dynamics of how these optimals evolve bear no resemblance to fully-developed
turbulence, which confirms the expectation discussed at the beginning of
\cref{sec:results} that in a real flow, optimals have to evolve on much shorter
time scales.
Thus, we can conclude that time scales between about \(T=0.7 h/\utau\) and
\(T=2.1 h/\utau\), and initial energy roughly between
\(e_{0}/E_{0}=\expnumber[3]{-5}\) and \(e_{0}/E_{0}=\expnumber{-4}\) give rise
to nonlinear optimals whose dynamics share distinctive features with
fully-developed turbulence.

\begin{figure}
  \centering
  \includegraphics[width=1.0\textwidth]{ 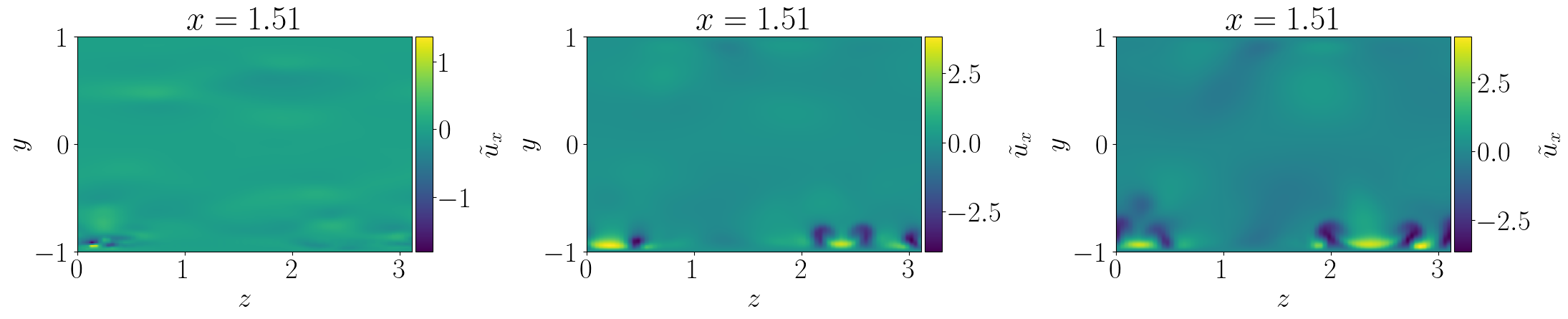 }
  \caption{Early part of the streamwise velocity evolution
    (\(t \utau/h = 0, 0.35, 0.7\) for comparison with \cref{fig:vel-2t0-7eminus5}) of the
nonlinear optimal (\(e_{0} / E_{0} = \expnumber[7.2]{-5}\)) for \(T \utau/h=2.1\)
plotted in \(y\)-\(z\)-plane.}
  \label{fig:vel-6t0-7eminus5}
\end{figure}

Qualitative visual inspection hints at a remarkable uniformity of all cases in
the nonlinear localised regime.
For example, comparing the nonlinear optimal at \(T=0.7 h/\utau\) and
\(e_{0}/E_{0} = \expnumber[7.2]{-5}\) (\cref{fig:vel-2t0-7eminus5}) with the one
found at \(T=2.1 h/\utau\) and \(e_{0}/E_{0} = \expnumber{-4}\)
(\cref{fig:vel-6t0-7eminus5}),
even though the localisation is a bit diminished in the latter one due to the
less restrictive initial energy, the general dynamics as well as length and time
scales are very similar.

\begin{figure}
  \centering
  \includegraphics[width=0.9\textwidth]{ 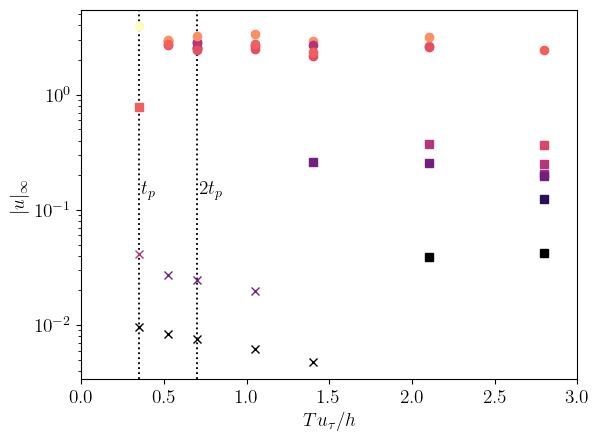 }
  \caption{Infinity norm for each case in \cref{fig:te0paramspace} with colours
    indicating $e_0$ (see \cref{fig:te0paramspace}).
    Crosses indicate the quasilinear regime,
    squares the nonlinear non-localised regime and circles the nonlinear localised
    regime.
    Despite different initial disturbance energies, all optimals
    of the nonlinear localised regime exhibit infinity norms in a very narrow
    range, which is not true for the optimals of the other regimes.}
  \label{fig:tinfnormparamspace}
\end{figure}

In the following, we pursue a more quantitative analysis to confirm the two
observations made above, i.e.\ that (i) the optimals in the nonlinear localised
regime show very similar dynamics to those seen in real turbulence, and that
(ii) this is true for all optimals in the nonlinear localised regime, even across the wide range of
\(T\) and \(e_{0}\) spanned by this regime.
Plotting the infinity norm of the initial disturbance velocity
$|\upert(\xv, 0)|_{\infty}$  (\cref{fig:tinfnormparamspace}) shows a clear
clustering of the different regimes.
One might expect $|\upert(\xv, 0)|_{\infty}$ to mainly depend on the initial
energy \(e_{0}\), because it determines the general magnitude of the velocity
disturbance. However, especially the cases that fall into the nonlinear
localised regime exhibit a remarkable uniformity in their respective
$|\upert(\xv, 0)|_{\infty}$.
This also confirms the qualitative observation in
\cref{fig:comp-nonlinear-optimals} that in the nonlinear localised regime,
additional initial energy does not lead to higher initial disturbance
amplitudes but mainly to reduced localisation.
Furthermore, it is reassuring that while the \(e_{0}/E_{0}\)-values are much
lower than typical disturbance energies in real turbulent flow, their
$|\upert(\xv, 0)|_{\infty}$-values are in the order of
\(10\) to \(20 \%\) of the mean centreline velocity, which is a much more
plausible value.

\begin{figure}
\begin{subfigure}{0.9\textwidth}
  \centering
  \includegraphics[width=0.7\textwidth]{ 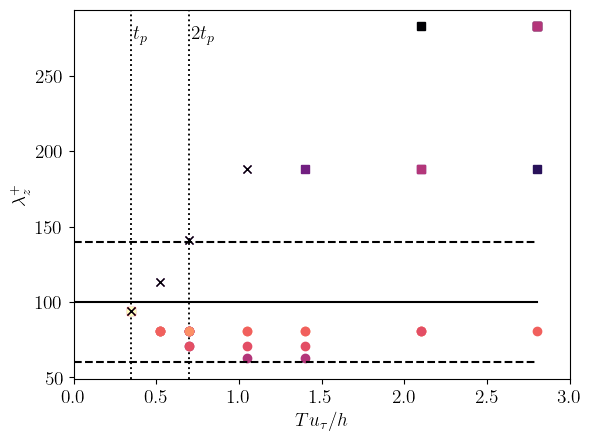 }
  \caption{ Streak spacing for each case in
    \cref{fig:te0paramspace} at \(t=0.3 \utau / h\) with colours indicating
    $e_0$ (see \cref{fig:te0paramspace}).
    Crosses indicate the quasilinear regime,
    squares the nonlinear non-localised regime and circles the nonlinear localised
    regime.
    Solid and dashed lines
respectively represent mean values and one standard deviation according to
\citet{butler1993}.}
  \label{fig:tlambdazparamspace}
\end{subfigure}
\\
\begin{subfigure}{0.9\textwidth}
  \centering
  \includegraphics[width=0.7\textwidth]{ 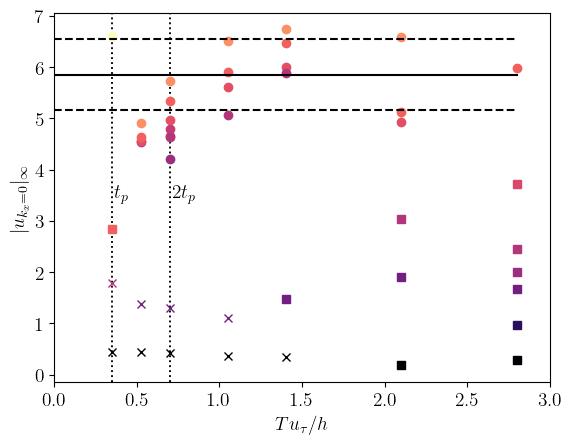 }
  \caption{Infinity norm  of the streak field at \(t=0.8 T\) with colours indicating
    $e_0$ (see \cref{fig:te0paramspace}).
    Crosses indicate the quasilinear regime,
    squares the nonlinear non-localised regime and circles the nonlinear localised
    regime.
    Solid and dashed lines
respectively represent mean values and one standard deviation taken from a DNS.}
  \label{fig:tstreakampparamspace}
\end{subfigure}
\caption{Comparison of streak length scales with DNS. Only the cases of the
nonlinear localised regime agree well with DNS, both in terms of streak spacing
(a) and streak amplitude (b).}
\end{figure}

This uniformity found in the nonlinear localised regime also manifests itself in
the streak spacing \(\lambda^{+}_{z}\), which is shown in
\cref{fig:tlambdazparamspace}.
Here, the streak spacing is obtained automatically by computing the
dominant \((k_{x} = 0, k_{z})\)-mode of the \(x\)-velocity at \(t=0.3 \utau / h\),
and then inferring the streak spacing from the \(k_{z}\)-value.
This streak spacing detection works very robustly for the quasilinear optimals,
but for the nonlinear optimals, the initial state is localised and does not
contain any streaks to speak of.  Towards the end of some of the runs, the
streaks break down or diminish in importance. As a result, the automatic streak
spacing detection is unreliable for early and late times of these runs.
However, in a fairly long intermediate phase from around
\(0.1 h/\utau\) to \(0.6 h/\utau\), when streaks stand out
distinctly, the method works well, which motivates the choice of
\(t=0.3 h/\utau\) as the reference time.
The result is that all runs that belong to the group of nonlinear localised
optimals show streak spacing values inside the expected range, which is only
true for some but not all the runs in the other groups.
In particular, it is apparent that linear optimals require a much more precise
choice of the time horizon \(T\) if they are to produce the correct streak
spacing, and that the optimals belonging to the nonlinear non-localised regime exhibit
unrealistically high streak spacing.
Further evidence for the uniformity and realism of the nonlinear localised
optimals is provided in \cref{fig:tstreakampparamspace}, which shows the
infinity norm of the streaks, i.e.\ the \(k_{x}=0\) part of the \(x\)-velocity
fields at
\(t=0.8 T\)
for each case.
This relatively late time is chosen here to allow the streaks enough time to not
only form but also grow.
Note that we consider the infinity norm rather than, say, the energy norm
here because the former is less sensitive to localisation, thus allowing for a
better comparison.
Unlike streak spacing, this metric is not widely discussed in the literature, so
that values for the mean and standard deviation of the infinity norm of streaks
were
obtained from 2000 DNS snapshots.
Again, only the optimals belonging to the nonlinear localised
regime produce streaks with a realistic infinity norm, 
though it is apparent that for the cases with relatively short time horizons,
the streaks are still a bit small as they have not had the time to fully establish
themselves yet.
One might object that the comparison is somewhat unfair, because the cases in
the quasilinear regime simply lack the initial energy to generate realistic
streak amplitudes, however, in the nonlinear non-localised regime, even when there is
sufficient initial energy, the streaks that form are not as pronounced and,
thus, do not seem as crucial to the overall evolution dynamics.

Another interesting point is that the general shape of these optimals closely
resembles that of optimal perturbations found in the transition problem. In the
parameter study by \citet{farano2016}, both the linear regime and the
nonlinear localised regime, which they refer to as highly nonlinear regime, seem
to have counterparts that can be observed in the transition problem as well,
although the time horizons considered in that study are much longer and the
initial energies smaller.

These results are reassuring and at the same time highlight the usefulness of
considering nonlinear optimals: whereas linear optimals require a very specific
time horizon \(T\) in order for their predicted steak spacing to agree with
fully-developed turbulence, nonlinear optimals are much more robust in this
regard, as every optimal in the nonlinear localised regime, which spans a
considerable range of \(T\) and \(e_{0}\), yields dynamics consistent with
fully-developed turbulence, despite the relatively simple equation system
employed here.

\subsection{Short time high energy optimals}
\label{sec:short-time-high}

\begin{figure}
  \centering
  \begin{subfigure}[t][\arraycolsep][t]{1.0\linewidth}
  \includegraphics[width=1.0\textwidth]{ 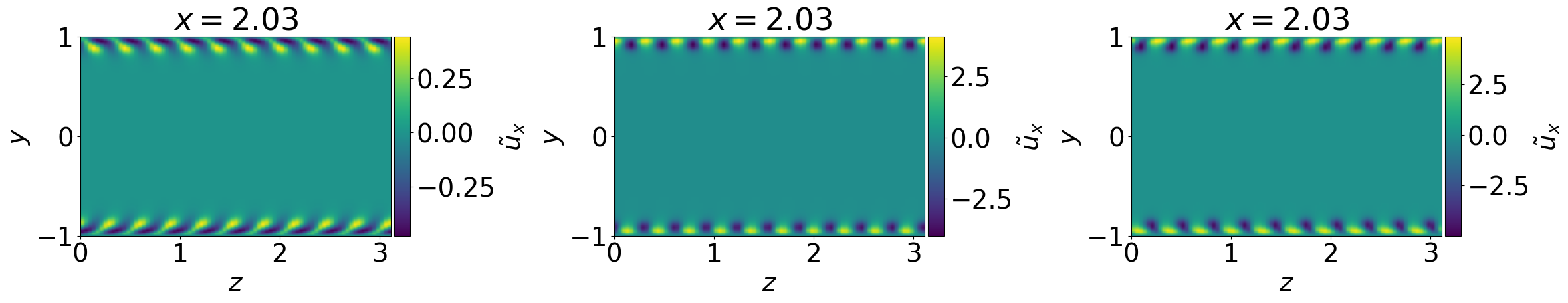 }
  \end{subfigure} \\
  \begin{subfigure}[t][\arraycolsep][t]{1.0\linewidth}
  \includegraphics[width=1.0\textwidth]{ 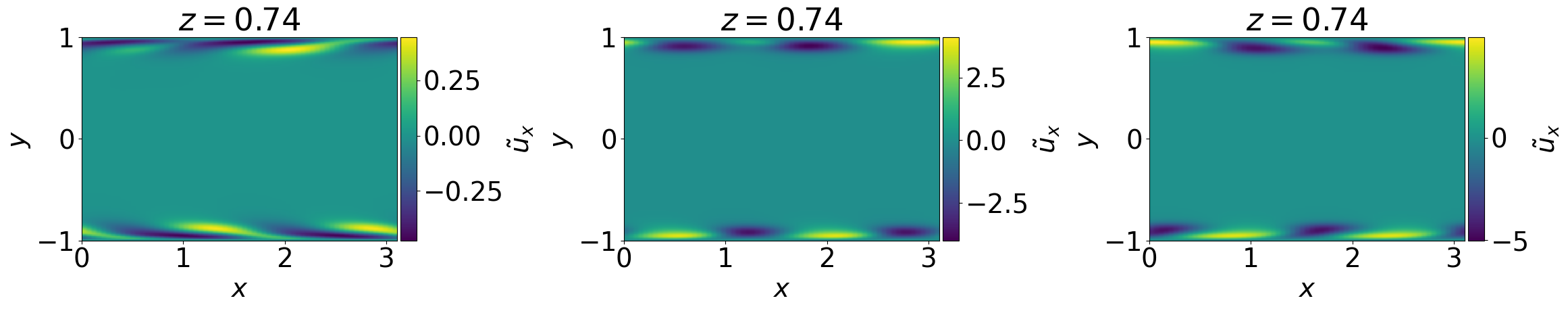 }
  \end{subfigure} \\
  \caption{Streamwise velocity evolution (\(t \utau/h = 0, 0.05, 0.1\)) of the
    nonlinear optimal (\(e_{0} / E_{0} = \expnumber[4]{-4}\)) for
    \(T \utau/h=0.1\) plotted in \(x\)-\(z\)-plane (top row) and in the \(x\)-\(y\)-plane (bottom row).}
  \label{fig:vel-0.3t0-1eminus3_xy_plane_short}
\end{figure}

Having expanded the parameter space to very long times, we now also address the
question of what happens for very short time horizons. Here, we choose
\(T = 0.1 \utau / h\). In order to compensate for this short time horizon, we
select a relatively large initial energy of \(e_{0}/E_{0} = \expnumber[4]{-4}\).
The evolution of this optimal, which grows by about an order of magnitude in
the given time horizon, is shown in \cref{fig:vel-0.3t0-1eminus3_xy_plane_short}.

Interestingly, there is no localisation in \(x\) or \(z\), and although the
disturbance is concentrated in \(y\)-direction to where the base shear is high,
it appears on both walls, again indicating that localisation is not important
for these parameters. This is partly expected due to the relatively high initial
energy, but the more important point is probably the short time horizon.
As can be inferred from the tilting of the disturbances visible in the bottom
row of \cref{fig:vel-0.3t0-1eminus3_xy_plane_short}, only the Orr mechanism
is important here, as there is no time for streaks to develop and make use of
the lift-up mechanism.
This is expected and consistent with the time scales observed in the temporal
evolution of the nonlinear optimals discussed in
\cref{sec:inner-time-scale,sec:expl-time-energy}.
The fact that there is only a single linear mechanism at play due to the short
time horizon thus helps explain why localisation is not observed.

\section{Conclusion}
\label{sec:conclusion}

In this work, we have investigated the mechanisms by which energy is transferred
from the mean to fluctuations in turbulent flow by  computing the nonlinear
optimals for various initial energies and time horizons.
The results show that for a large range of intermediate time horizons, given
initial energies large enough to trigger nonlinear effects, localised nonlinear
optimals with very similar dynamics emerge. Important aspects of these dynamics,
such as the streak spacing and amplitudes, are in good agreement with
fully-developed turbulence.
Interestingly, the lower end of this range of time horizons is consistent with
previous estimates of the time scales on which primary and secondary linear
processes operate.

These results confirm that the evolution of nonlinear optimals reflects crucial
aspects of turbulent dynamics, despite the simplicity of our model. At the same
time, the optimisation serves to remove any non-essential noise not contributing
to disturbance growth, which would be difficult to filter out in other methods
such as DNS.
This makes nonlinear optimals an attractive concept for isolating the important
dynamics, allowing one to study turbulence in a ``clean'' environment.
At the same time, the present analysis provides upper bounds for the level of
energy transfer to the turbulent fluctuations, which could serve as a foundation
for future work aimed at better understanding under which circumstances
turbulence is sustained.
When compared to computationally less expensive linear methods, two major
advantages of this approach are (i) the conceptual simplicity -- in particular
there is no need to prescribe a streak amplitude as is the case for methods
coupling linear processes ``manually'' \citep[e.g.][]{kerswell2022,ciola2024} --
and (ii) the robustness with which correct values for important characteristics
of the dynamics, namely streak spacing and amplitude, are found with respect to
parameter variations.

Future work could be aimed at combining the method of nonlinear optimisation
with numerical experiments, in which equations are deliberately altered to test
the effect. In the past, such numerical experiments have been done using DNS,
however, this has the limitation that one relies on DNS snapshots of realistic
turbulence as initial conditions to make predictions about an artificial
setting. This may make conclusions less reliable compared to using nonlinear
optimisation, which are fully self-contained.

Furthermore, the role of the base profile should be investigated. Understanding
how different base profile affect disturbance growth could help answer the
question of why a particular base profile (i.e.\ the mean) is selected in reality
instead of some different profile.
This line of reasoning could ultimately even provide an avenue towards
predicting the mean profile in shear flows at moderate Reynolds number if a
corrected version of \posscite{malkus1956} theory could be formulated.

\noindent
\textbf{Declaration of interests:}
The authors report no conflict of interest.

\noindent
\textbf{Acknowledgements:}
We thank Sergio Hoyas and Patrick Keuchel for fruitful discussions about the
solver implementation.

\noindent
\textbf{Funding:}
The work leading to this publication was supported by the PRIME programme of the
German Academic Exchange Service (DAAD) with funds from the German Ministry of
Education and Research (BMBF).

\noindent
\textbf{Data availability statement:}
The data that support the findings of this study are openly available in Pangaea at https://doi.pangaea.de/10.1594/PANGAEA.983358.
\bibliographystyle{jfm}
\bibliography{bibliography}

\appendix
\section{}\label{appA}
\subsection{Solver validation}
\label{sec:solver-validation}

\begin{figure}
  \centering
  \includegraphics[width=0.7\textwidth]{ 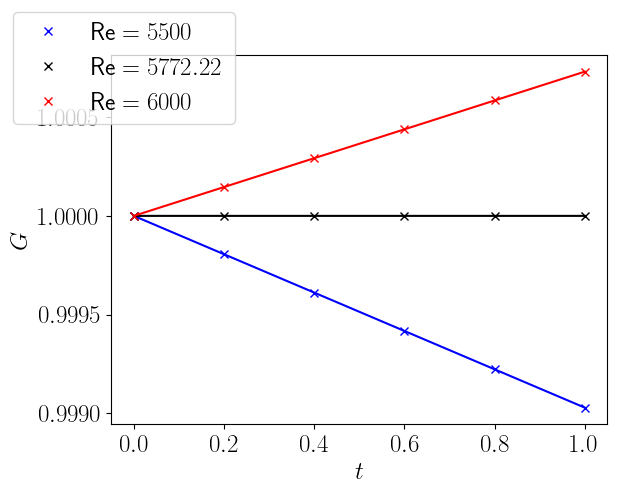 }
  \caption{Growth rates of the most unstable mode at $\text{Re}=5500, 5772.22, 6000$,
    with \(\alpha=1.02056; \beta=0\):
    Analytic values (lines) compared with numerical results (crosses)}
  \label{fig:growthrates}
\end{figure}

\begin{figure}
  \centering
  \includegraphics[width=0.7\textwidth]{ 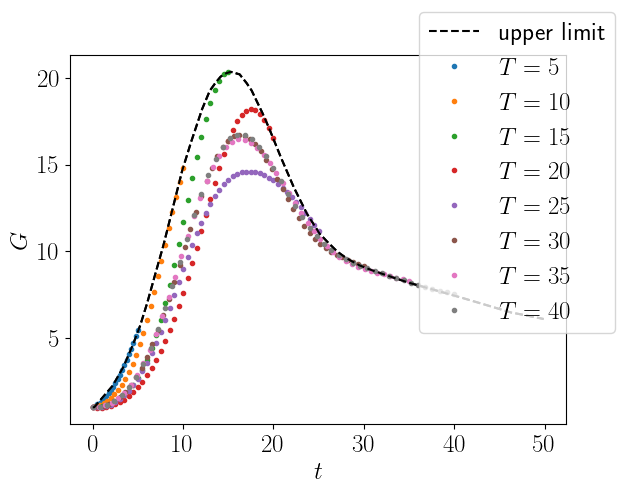 }
  \caption{Growth rates of the linear optimal at $\text{Re}=3000$ with
\(\alpha=1, \beta=0\) for different values of the time horizon \(T\), compared
with the optimal growth plotted over time \citep[dashed line;
from][]{reddy1993}.}
  \label{fig:transientgrowth}
\end{figure}

The solver presented in \cref{sec:computational-method} was validated by
reproducing the growth rates of single modes (see
\cref{fig:growthrates}), and by testing the transient growth of linear optimals
(see \cref{fig:transientgrowth}).
Both the single modes and the linear optimals were computed by the
solver, which contains an implementation of the relevant linear stability
problem in the spirit of \textcite{reddy1993} for this specific purpose.
As can be seen in the two figures, the agreement is excellent.

\subsection{Influence of the domain size}
\label{sec:infl-doma-size}

\begin{figure}
  \centering
  \includegraphics[width=1.0\textwidth]{ 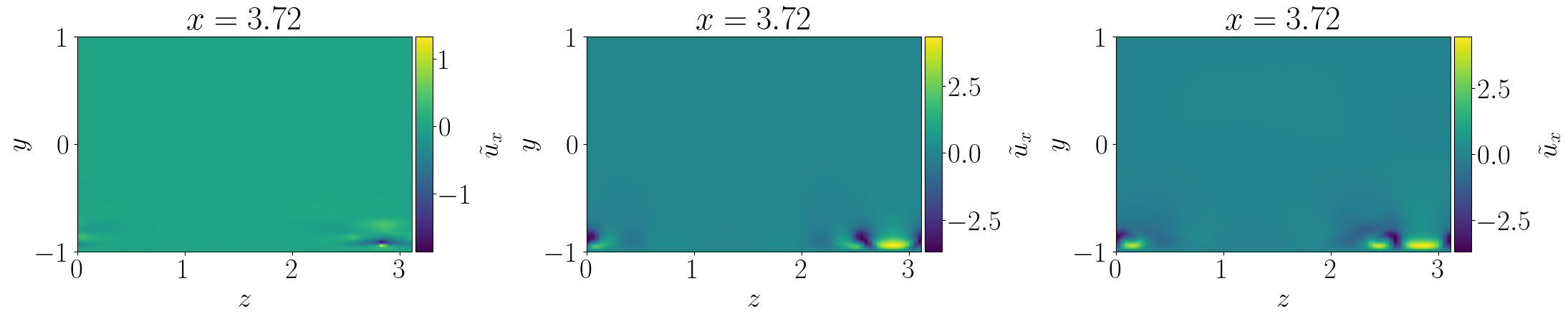 }
  \caption{Streamwise velocity evolution (\(t \utau/h = 0, 0.35, 0.7\)) of the optimal for \(T \utau/h=0.7\) and
    \(e_0/E_{0} = \expnumber[3]{-5}\) in a larger channel.}
  \label{fig:vel-2t0-3eminus5-large}
\end{figure}

\begin{figure}
  \centering
  \includegraphics[width=1.0\textwidth]{ 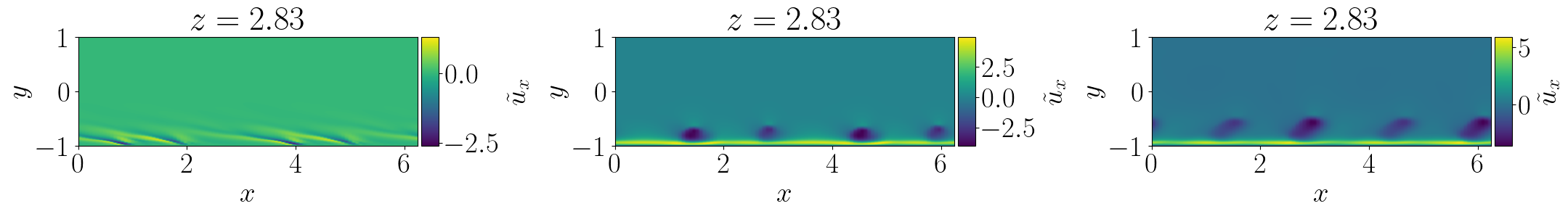 }
  \caption{Streamwise velocity evolution (\(t \utau/h = 0, 0.35, 0.7\)) of the optimal for \(T \utau/h=0.7\) and
    \(e_0/E_{0} = \expnumber[3]{-5}\) in a larger channel plotted in the \(x\)-\(y\)-plane.}
  \label{fig:vel-2t0-3eminus5-xy-large}
\end{figure}

To justify the choice of the fairly small domain used in this study, we compare
it to the more commonly used domain size of \(2\pi \times 2 \times \pi\), for which a
resolution of \(96 \times 129 \times 80\) is used.
\cref{fig:vel-2t0-3eminus5-large} shows the evolution of the optimal at
\(T \utau/h = 0.7\) and \(e_{0}/E_{0}=\expnumber[3]{-5}\) on this larger domain.
As can be seen by comparison with the left column in
\cref{fig:comp-nonlinear-optimals}, the dynamics are essentially indistinguishable.
The gain of \(33.63\) is also very close to the small channel value of \(34.82\).
Interestingly, as is revealed by comparing \cref{fig:vel-2t0-3eminus5-xy-large}
with \cref{fig:vel-2t0-7eminus5_xy_plane} (the slightly higher \(e_{0}\) has no
significant effect on the dynamics), four instead of two clusters are present in
the initial condition, and the streamwise length scale of the low-speed streaks
is also the same as four instead of two low-speed regions can be seen in this
plane.

\begin{figure}
  \centering
  \includegraphics[width=1.0\textwidth]{ 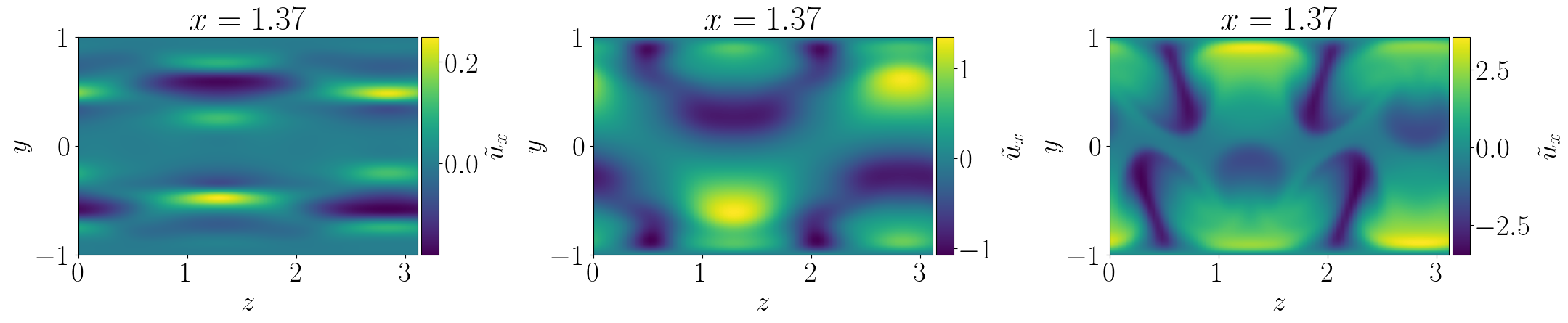 }
  \caption{Streamwise velocity evolution (\(t \utau/h = 0, 1.4, 2.8\)) of the optimal for \(T \utau/h=2.8\) and
    \(e_0/E_{0} = \expnumber[3]{-5}\) in a larger channel.}
  \label{fig:vel-8t0-3eminus5-large}
\end{figure}

A similar comparison was also done for the very different time horizon of
\(T=2.8\), again at \(e_{0}/E_{0} = \expnumber[3]{-5}\), placing the
corresponding optimal into the nonlinear non-localised regime. Again, the quantitative
agreement of the gain, which was calculated to be \(360.85\) is in excellent
agreement with the small channel value of \(361.20\). Furthermore, comparison of
\cref{fig:vel-8t0-3eminus5-large} with \cref{fig:vel-8t0-3eminus5} reveals good
qualitative agreement of the time evolution of the optimals for each channel size.

We can thus conclude that the smaller channel of \(\pi \times 2 \times \pi\) is
sufficient for the purposes of this investigation.

\subsection{Gradient-based optimisation algorithm}
\label{sec:grad-desc-algor}

We now turn to a more detailed description of the optimisation approach roughly
introduced in \cref{sec:computational-method}.
Generally speaking, any gradient-based optimisation algorithm has two components: First,
selecting a step direction, and, second, selecting a step size determining
how far to move along this direction.
Note that in the literature, optimisation problems are often phrased as minimisation
problems. Here, since we are trying to maximise the energy gain, we use language
to reflect that we are solving a maximisation problem. The mathematics are
obviously completely equivalent up to flipping signs. Notable, we refer to
gradient ascent rather than gradient descent.
A natural choice for the ascent direction is the steepest direction (given by
the gradient), however, in the present work, we follow \textcite{cherubini2011}
and use the conjugate gradient algorithm with the Polak-Ribière formula \parencite{polak1969}.
In concrete terms, the ascent direction \(\vv{p}^{n+1}\) is determined using
\begin{equation}
  \label{eq:3}
  \vv{p}^{n+1} = \vd{\lagrangian}{\upert^{n}(\xv, 0)} + \beta^{n} \vv{p}^{n} = \lambda^{n} \upert^{n}(\xv, 0) - \uadj^{n}(\xv, 0) + \beta^{n} \vv{p}^{n}.
\end{equation}
At the first step, \(\beta^{0}\) is initialised with zero. Afterwards, it is calculated
by
\begin{equation}
  \label{eq:4}
  \beta^{n+1} = \frac{\transpose{(\vv{g}^{n+1})} ( \vv{g}^{n+1} - \vv{g}^{n} )}{\transpose{(\vv{g}^{n})} \vv{g}^{n}},
\end{equation}
where \(\vv{g}\) is the steepest ascent direction.
Note that some additional complexity arises from the fact that
\cref{eq:lagrgrad} depends on the Lagrange multiplier \(\lambda\), which ensures the
initial energy constraint. At step \(n\), for a given step size \(\alpha^{n}\) and
conjugate parameter \(\beta^{n}\) , \(\lambda^{n}\) is determined implicitly from the
formula
\begin{equation}
  \label{eq:2}
  \frac{1}{2V} \volint{\norm{\upert^{n+1}}^{2}} = e_{0},
\end{equation}
where the new guess \(\upert^{n+1}\) is given by
\begin{equation}
  \label{eq:1}
  \upert^{n+1} = \upert^{n} + \alpha \vv{p}^{n},
\end{equation}
and the ascent direction \(\vv{p}^{n}\) is obtained from \cref{eq:3}.
Here, we solve \cref{eq:2} for \(\lambda\) using a standard Newton method, which
the necessary gradients being obtained by automatic differentiation.

Unlike \textcite{cherubini2011}, who adjust the step size depending on whether
the last iteration was successful in increasing the objective function, our
solver includes the option of using a line search. This can save time under some
circumstances, but may also be counterproductive due to its inability to
``escape'' local optima, so while line search was used most of the time in
obtaining the results presented here, some calculations fully or partially
relied on the simpler adaptive step method described in
\textcite{cherubini2011}.
The crucial advantage of line search algorithms are that instead of simply
accepting an optimisation step, it is first checked if taking this step actually
leads to a sufficient increase in the objective function, and only then is the
step accepted. Thus, line search ensures that the objective function increases
in each iteration.
Moreover, even though more function evaluations are necessary in each time step,
the step size is adaptively increased if possible, leading to fewer iterations
until convergence. Whether this trade-off makes sense depends on the problem, but
usually and also in the present case, it does.

The general idea behind a line search algorithm is to maximise the objective
function along the chosen ascent direction. However, in practice, it is not
a good use of computational resources to find an exact optimum, but it is better
to instead settle on a step size that yields a reasonable improvement and is
easy to find. Depending on the problem, many approaches exist for this. In the
present work, once an ascent direction is determined, a step using the current
step size \(\alpha_{i}\) is taken, and the function is evaluated at the new guess.
If the increase is big enough, i.e.\ fulfils the Armijo condition \parencite{armijo1966}
\begin{equation}
  \label{eq:armijo}
  f(x + \alpha_{i} p) - f(x) \geq \alpha_{j} t,
\end{equation}
where \(f\) is the objective function and \(t\) is a parameter between \(0\) and
\(1\), the new guess is either accepted directly, or \(\alpha_{i}\) is increased
until \labelcref{eq:armijo} is no
longer satisfied. Note that trying to increase the step size requires additional
function evaluations, so in practice, it should usually not be done in every
iteration, but only, say, every third iteration. However, not trying to increase
the step size once in a while will likely lead to slow convergence due to a
possibly too small step size.
If \labelcref{eq:armijo} is not satisfied, \(\alpha_{i}\) is decreased until it is.
The step size for the new iteration \(\alpha_{i+1}\) is then chosen to be the final
step size of the old iteration.
Note that upper and lower limits for the step size \(\alpha_{i}\) can be imposed to
ensure reasonable behaviour of the solver. This was done in the present work,
though the borders were chosen quite liberally as a restriction of the steps
size should in principle not be required.
The convergence criterion is that the relative change of the objective function between
two successive iterations lies below a critical value (\(\expnumber{-8}\) is
chosen in the present work).

\end{document}